\documentclass[12pt]{article}

\usepackage[dvips]{graphicx}

\newcommand{\rig}{\rightarrow}

\renewcommand{\d}{{\mathrm{d}}}
\newcommand{\be}{\begin{eqnarray*}}
\newcommand{\ee}{\end{eqnarray*}}
\newcommand{\gl}[1]{(\ref{#1})}
\newcommand{\ta}[2]{ \frac{ {\mathrm{d}} #1 } {{\mathrm{d}} #2}}

\newcommand{\bee}{\begin{eqnarray}}
\newcommand{\eee}{\end{eqnarray}}

\newcommand{\mr}{\mathrm}

\def\lesssim{\mathrel{\raisebox{-.6ex}{$\stackrel{\textstyle<}{\sim}$}}}

\def\slashed#1{#1\llap{\sl/}}

\begin{document}
\begin{titlepage} 
\nopagebreak  
{\flushright{ 
        \begin{minipage}{5cm}
         KA--TP--32--2008  \\       
         SFB/CPP-08-103  \\
%        {\tt hep-ph/08xxxxx}\hfill \\ 
        \end{minipage}        } 
} 
\vfill 
\begin{center} 
{\LARGE \bf  
QCD Corrections to \\[2mm] 
Vector-Boson Fusion Processes \\[5mm]
in Warped Higgsless Models 
} 
\end{center}
\vfill
\begin{center} {\large  \bf 
C.~Englert$^{a,}$\footnote[1]{Electronic address: englert@particle.uni-karlsruhe.de}, B. J\"ager$^{b,}$\footnote[2]{Electronic address: bjaeger@physik.uni-wuerzburg.de} and D.~Zeppenfeld$^{a,}$\footnote[3]{Electronic address: dieter@particle.uni-karlsruhe.de}
}
\end{center}

\vskip 0.5cm

\begin{center}   
\vskip .2cm  
$^{a}$ Institut f\"ur Theoretische Physik, 
Universit\"at Karlsruhe, 76128 Karlsruhe, Germany\\
\noindent
$^{b}$ Institut f\"ur Theoretische Physik und Astrophysik, Universit\"at W\"urzburg, 97074~W\"urzburg, Germany\\ 
 
 \vskip 
1.3cm     
\end{center} 
 
\nopagebreak 
\begin{abstract}
We discuss the signatures of a representative Higgsless model with ideal fermion delocalization in vector-boson fusion processes, focusing on the gold- and silver-plated decay modes of the gauge bosons at the CERN-Large Hadron Collider. For this purpose, we have developed a fully-flexible parton-level Monte-Carlo program, which allows for the calculation of cross sections and kinematic distributions within experimentally feasible selection cuts at NLO-QCD accuracy. We find that Kaluza-Klein resonances give rise to very distinctive distributions of the decay leptons. Similar to the Standard Model case, within the Higgsless scenario the perturbative treatment of the vector-boson scattering processes is under excellent control. 
\end{abstract} 
\vfill 
\end{titlepage} 
\newpage               
%
%--------------------------------------------------------------------
%
%--------------------------------------------------------------------
%			MAIN SECTION
%--------------------------------------------------------------------
%
\section{Introduction}
The mechanism of electroweak symmetry breaking remains one of the most important issues to be addressed at the CERN Large Hadron Collider (LHC). 
Even though the Standard Model (SM) accounts extraordinarily well for many measurements, the postulated Higgs mechanism could not yet be verified experimentally. Moreover, the SM exhibits various theoretical caveats that have led to a plethora of alternative models of electroweak symmetry breaking (EWSB). One of the flaws of the SM that is considered to be seminal to a more fundamental theory is the so-called ``hierarchy problem'', which is tantamount to the radiative instability of the scalar Higgs boson's vacuum expectation value $v$ in the SM. The interpretation of the observed hierarchy $v\ll M_{\mathrm{Pl}}$ between the scale of EWSB and the four-dimensional Planck scale $M^{d=4}_\mr{Pl}$ in terms of additional, compactified dimensions has lead to the renaissance of theories with extra compactified dimensions \cite{Kaluza:1921tu,Arkani-Hamed:1998rs,Randall:1999ee}. In particular, the Randall-Sundrum RSI scenario equipped with a bulk gauge group $\mr{SU}(2)_L\times \mr{SU}(2)_R \times \mr{U}(1)_{B-L}$, along with its dual interpretation driven by the AdS/CFT conjecture \cite{Maldacena:1997re} allows for consistent gauge symmetry breaking by boundary conditions, without relying on scalar relics in the particle spectrum, yet being consistent with electroweak data \cite{Csaki:2003dt, Agashe:2003zs} to leading order. These minimal models of EWSB suffer,
however, from a tension between high-scale partial wave unitarity and consistency with precision electroweak data \cite{Davoudiasl:2003me, Barbieri:2003pr, Barbieri:2008cc}. 

Crucial to the consistency with measurements of the oblique corrections to leading order, that is mainly the, typically $\mathcal{O}(1)$, positive size of Technicolor-like theories'  $S$ parameters, is that light fermions added to the theory have to be allowed to propagate into the bulk with small to vanishing couplings to the non-SM states of the Kaluza-Klein (KK) gauge tower \cite{Cacciapaglia:2004rb}. This leads to a suppressed production of Kaluza-Klein states by Drell-Yan processes, and strongly motivates vector-boson fusion (VBF) reactions as possible discovery channels for the non-SM resonances. Because of the dimensionful couplings of the higher-dimensional gauge theory,  the model is valid only up to a scale set by Naive Dimensional Analysis \cite{Papucci:2004ip}.

The compactification of the additional dimension in bulk-gauged Randall-Sundrum models results in an infinite tower of neutral as well as charged weak boson states, due to the quantization of momenta along the compactified dimension. This constitutes a phenomenologically striking implication of the scenario: The appearance of spin-one resonances in VBF processes with three-leptons-plus-missing transverse momentum $\slashed{p}_T$ in the final state is in sharp contrast to the SM as well as to general two-Higgs doublet models, where resonances are heavily suppressed by the absence of charged Higgs couplings to weak bosons and by the smallness of the couplings of (pseudo-)scalar Higgs bosons to the external quarks. Together with massive excitations of the neutral gauge bosons, this leads to interesting phenomenological implications at the LHC, which have been studied in \cite{Birkedal:2004au,Englert:2008tn}.
 
Not addressing the pending model-building provisos, we consider this minimal model as a prototype of a perturbatively calculable scenario of dynamical EWSB with iso-vector resonances. This allows us to generically access the phenomenological implications of composite vector excitations, and to examine the impact of QCD corrections on their production in VBF. 
This is straightforwardly done by employing the parton-level Monte Carlo
program \textsc{Vbfnlo} \cite{vbfnlo}, which features the calculation of cross sections and differential distributions of VBF processes at NLO-QCD accuracy. Due to the modular implementation of the QCD corrections, \textsc{Vbfnlo} can be easily adapted for the study of the VBF phenomenology of non-standard models of EWSB. The present study provides an example for such an implementation
of non-SM effects.

The article is organized as follows: The theoretical aspects of the Warped Higgsless model relevant for our analysis are sketched in Sec.~\ref{sec:hless}. For a more thorough review, we refer to the literature, e.~g. Ref.~\cite{Davoudiasl:2003me}.
We discuss the mass spectra and couplings used for our numerical study, justify the approximations made, and give details on the implementation of the Kaluza-Klein excitations into \textsc{Vbfnlo}.  
In Sec.~\ref{sec:lhcpred}, tree-level results are presented for the production modes $W^+W^- jj$, $W^\pm Z jj$, and $ZZjj$, including off-shell and finite width effects of the gauge boson decays for different Kaluza-Klein mass scenarios.  In Sec.~\ref{sec:qcdcor} we discuss the impact of the dominant NLO-QCD corrections on the phenomenology of the considered Higgsless model. More details on the model can be found in the Appendix.
%
%--------------------------------------------------------------------
%			MAIN SECTION
%--------------------------------------------------------------------
%
\section{Warped Higgsless Kaluza-Klein scenario}
\label{sec:hless}
In this work, we consider a Warped Higgsless model which is a $\mr{SU(2)}_R\times \mr{SU(2)}_L \times \mr{U(1)}_X$ bulk-gauged version of the RSI scenario of Ref.~\cite{Randall:1999ee}. In this model, the gauge theory is defined on a slice of a five-dimensional Anti-de Sitter space (AdS$_5$) with metric
\bee
\label{rsmetric}
\d s^2 = {R^2\over y^2} \left( g_{\mu\nu} \d x^\mu \d x^\nu - \d y^2 \right)\,, 
\eee
where $x$ denotes the coordinate of the ordinary four dimensions and $y$ the coordinate along the extra dimension. 
Neglecting heavy-flavor contributions, $X$ can be identified as  baryon-minus-lepton number, $B-L$. 
We focus on a scenario with identical couplings for the $\mr{SU(2)}_L$ and $\mr{SU(2)}_R$ subgroups in five dimensions, $g_{5,L}=g_{5,R}=g_5$. The extra dimension is compactified on an interval, $R \leq y \leq R'$, and bounded by two branes, which are referred to as Planck (UV) and TeV (IR) brane, respectively. The symmetry breaking pattern of the scenario is depicted in Fig.~\ref{fig:hlessmod}: 
%
%
%%%%%%%%%%%%%%%%%%%%%%%%%%%%%%%%%%%%%%%%%%%%%%%%%%%%%%%
%
\begin{figure}[!t]
\begin{center}
\includegraphics[width=0.65\textwidth]{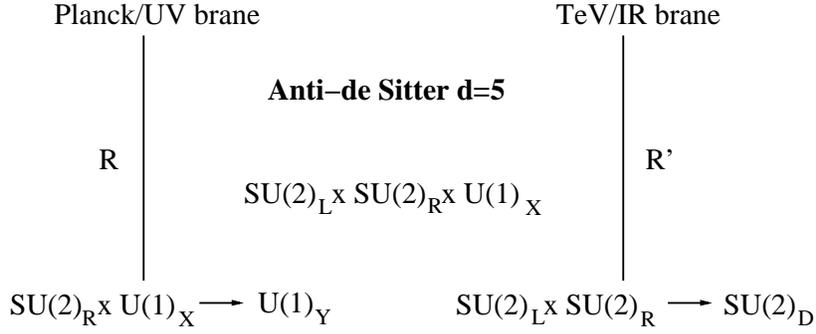}
\caption{\label{fig:hlessmod} \emph{Sketch of the Higgsless model's symmetry breaking pattern \cite{Csaki:2003dt}.}}
\end{center}
\vskip -0.4cm
\end{figure}
%
%%%%%%%%%%%%%%%%%%%%%%%%%%%%%%%%%%%%%%%%%%%%%%%%%%%%%%%
%
%
%
The $\mr{SU(2)}_R\times \mr{SU(2)}_L \times \mr{U}(1)_X$ gauge symmetry is broken on the Planck brane to the electroweak group $\mr{SU(2)}_L\times \mr{U}(1)_Y$. Electroweak symmetry breaking takes place on the TeV brane, reducing $\mr{SU(2)}_L\times \mr{SU(2)}_R$ to $\mr{SU(2)}_D$, where $D$ denotes the diagonal subgroup. 
Restricting $y$ to a finite interval, the gauge sector of the effective
four-dimensional theory contains a Kaluza-Klein tower of $W_k^\pm$ and
$Z_k$ bosons of mass $m_{W_k}$ and $m_{Z_k}$, respectively, and a
massless mode $Z_{k=0}$ to be identified with the photon. The first massive modes are labeled by $k=1$ and are interpreted as the $W^\pm$ and $Z$ bosons of the SM.

Relations between the couplings and masses of the Kaluza-Klein resonances and the SM gauge bosons can be derived from constraints on the high-energy behavior of longitudinal SM gauge boson scattering. In the absence of a Higgs boson, the corresponding SM scattering amplitudes grow with energy, thus violating unitarity at energies beyond about 1.2~TeV.  In Higgsless models, this deficiency is cured by the following sum rules on the various masses and the triple and quartic gauge boson couplings,   
\bee
\label{sumww1}
g_{W_1W_1W_1W_1}&=& \sum_{k\geq 0} g_{W_1W_1Z_k}^2\,,\\
\label{sumww2}
4m_{W_1}^2g_{W_1W_1W_1W_1} &=& 3\sum_{k\geq 1} m_{Z_k}^2 g_{W_1W_1Z_k}^2\,, \\
\label{sumwz1}
g_{W_1W_1Z_1Z_1}&=&\sum_{k\geq 1} g_{W_kW_1Z_1}^2\,,\\
\label{sumwz2}
2(m_{Z_1}^2+m_{W_1}^2)g_{W_1W_1Z_1Z_1}&=& \sum_{k\geq 1} g_{W_kW_1Z_1}^2 \left( 3m_{W_k}^2 - {(m_{Z_1}^2-m_{W_1}^2)^2\over m_{W_k}^2 }\right) \,.
\eee
These constraints follow from the orthogonality and completeness of the gauge bosons' wave functions along the $y$ direction \cite{Csaki:2003dt,Birkedal:2004au} and from requiring gauge invariance in five dimensions. 

Relations for the remaining gauge boson couplings are determined by the $\mr{U(1)}_\mr{QED}$ gauge symmetry and orthogonality of the Kaluza-Klein wave functions along the extra dimension.
%
%--------------------------------------------------------------------
%	 		SUBSECTION
%--------------------------------------------------------------------
%
\subsection{Mass spectrum of the Kaluza-Klein states}
Interpreting the $k=1$ excitations of the charged Kaluza-Klein tower as the $W^\pm$ bosons of the SM fixes the location of the IR-brane as a function of the $W_1^\pm$ mass and the UV-brane location. This leaves the $W_k^\pm$ mass spectrum entirely determined, once the UV brane location $R$ is fixed. The coupling of the unbroken gauge group's massless state $Z_0$ to these $W_1^\pm$ bosons is identified with the QED coupling
\be
g_{W_1W_1Z_0}= e\,.
\ee
We choose the gauge kinetic terms to be normalized canonically. This yields a relation between $g_5$  and the $\mr{U}(1)_X$ coupling $\tilde{g}_5$ (cf.~App.~\ref{app:whm}),
\be
e^2={ g_5^2 \tilde{g}_5^2 \over (2\tilde{g}_5^2+g_5^2) R \log (R'/R)}\,.
\ee
%
%
%%%%%%%%%%%%%%%%%%%%%%%%%%%%%%%%%%%%%%%%%%%%%%%%%%%%%%%
%
\begin{figure}[!ht]
\begin{center}
\includegraphics[height=7.5cm,width=7.5cm]{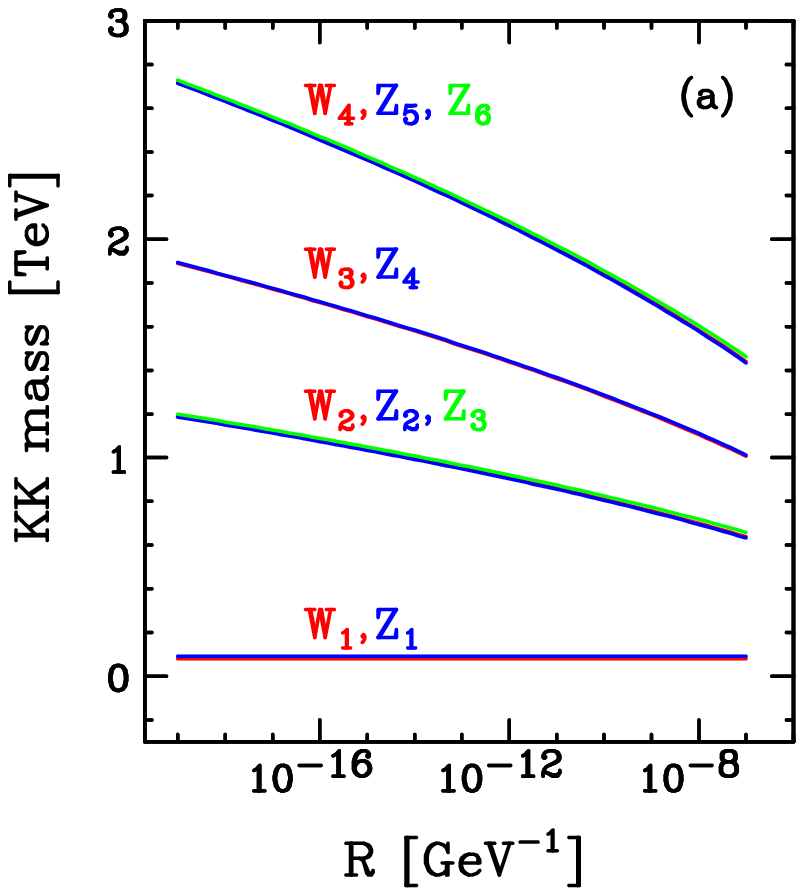}
\hfill
\includegraphics[width=7.5cm,height=7.5cm]{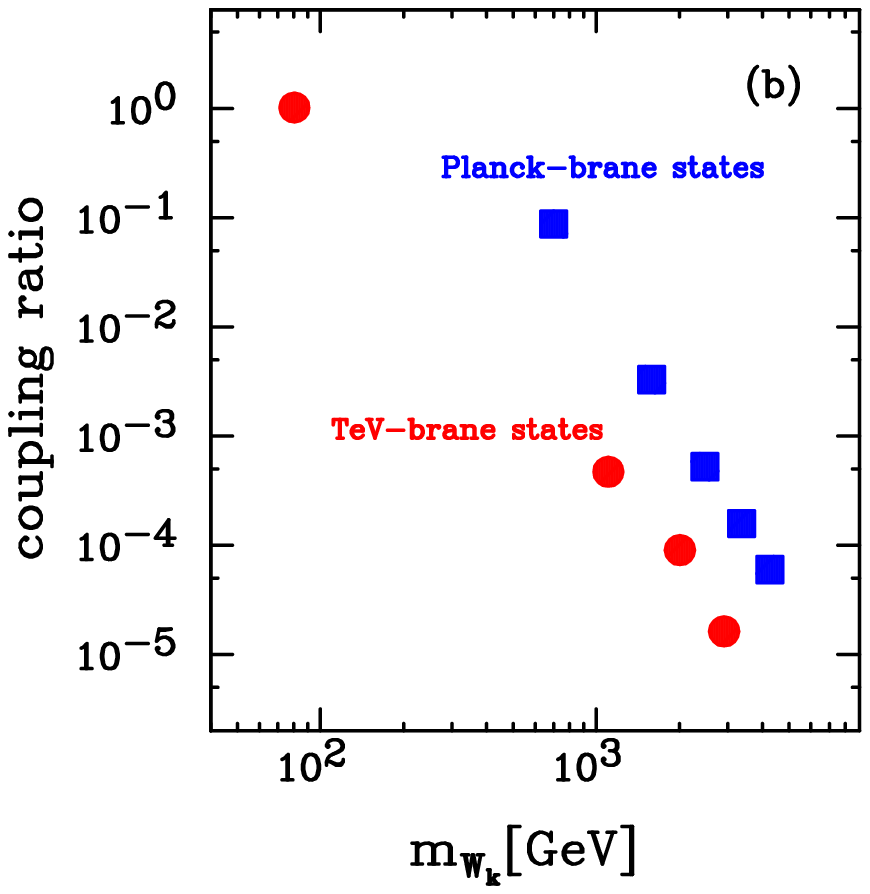}
\caption{\label{fig:wzmass}\emph{
Massive Kaluza-Klein $Z_k$ and $W_k$ states (a) as functions of the Planck brane localization $R$ and (b) 
couplings of the Kaluza-Klein $W_k^\pm$ states to the SM modes, normalized to the SM value of the $W^+W^-Z$ coupling, for a Kaluza-Klein spectrum with the mass of the lowest non-SM excitation being  $m_{W_2}=700~\textnormal{GeV}$. Circles indicate the TeV-brane related states, squares refer to the Planck-brane related states.}}
\end{center}
\vskip -0.4cm
\end{figure}
%
%%%%%%%%%%%%%%%%%%%%%%%%%%%%%%%%%%%%%%%%%%%%%%%%%%%%%%%
%
%
\noindent The mass of the SM $Z$ boson, identified with the first massive, neutral Kaluza-Klein state, eventually fixes the absolute values of the couplings $g_5$ and  $\tilde{g}_5$ as functions of $R$. Thus, the neutral gauge bosons' mass spectrum is fully determined as a function of $R$, see Fig.~\ref{fig:wzmass}. 

The mass spectrum exhibits characteristic features, due to the mixing of the gauge fields by the chosen set of boundary conditions. The states can be 
classified into two categories:
First, Kaluza-Klein states that arise mostly from the Planck-brane located symmetry breaking pattern along with a massive excitation of the photon, such as the states ($W_2^\pm,Z_2,Z_3$).  Second,  states following from TeV-brane located symmetry breaking analogous to the SM, like ($W_3^\pm,Z_4$). 
These states can be considered as Kaluza-Klein excitations of the Standard Model gauge bosons -- in contrast to states that follow from Planck-brane located symmetry breaking. For the latter, the weak $Z$ state is lighter than the corresponding $W$ state, so that no electroweak mixing angle analogon to the SM can be defined for these excitations.

The difference between Planck-brane and TeV-brane related Kaluza-Klein states is also visible in the structure of their couplings to the SM gauge bosons, sketched in Fig.~\ref{fig:wzmass}. 
The circles in Fig.~\ref{fig:wzmass} refer to the TeV-brane related Kaluza-Klein states, while the squares indicate the couplings of the Planck-brane excitations. Obviously, low excitations couple to the SM bosons more strongly than high-mass states, implying rapid convergence of the sum rules, Eqs.~\gl{sumww1}-\gl{sumwz2}.  
According to our choice of gauge couplings, the model is fixed by the single parameter $R$. Matching $g_{W_1W_1Z_0}=e$ results in an $R$-dependence of $g_{W_1W_1Z_1}$. The latter deviates from the SM value by less than $2\%$ in the considered parameter range $R\lesssim 10^{-7}~\textnormal{GeV}^{-1}$  \cite{Cacciapaglia:2004rb}, which is compatible with LEP2 data \cite{LEPEWWG}. 
%
%
%--------------------------------------------------------------------
%			SUBSECTION
%--------------------------------------------------------------------
%
\subsection{Implementation into the Monte Carlo program \textsc{Vbfnlo}}
The calculation of the Feynman diagrams contributing to the production of four leptons plus two jets via VBF in Warped Higgsless models is performed in complete analogy to the SM case presented in Ref.~\cite{Zeppenfeld:2007ur} and implemented in the parton-level Monte Carlo program \textsc{Vbfnlo} \cite{vbfnlo}. 
We consider 
$pp\to e^+\nu_e \mu^-\bar\nu_\mu jj$, 
$pp\to e^+ \nu_e \mu^+\mu^- jj$, and $pp\to e^-\bar\nu_e \mu^+\mu^- jj$ via VBF, which for simplicity are referred to as $W^+W^-jj$, $W^+Zjj$, and $W^-Zjj$ production, respectively, and VBF $ZZjj$ production with subsequent decay of the $Z$ bosons into four charged leptons, 
$pp\to e^+ e^- \mu^+\mu^- jj$, or into two charged leptons and two neutrinos, $pp\to e^+ e^- \nu_\mu\bar\nu_\mu jj$. 
The structure of the calculation is illustrated by means of the $W^+W^-jj$ mode in the following. The VBF $W^+Zjj$,  $W^-Zjj$, and  $ZZjj$ channels are tackled in a very similar manner. 

\begin{figure}[!ht]
\begin{center}
\includegraphics{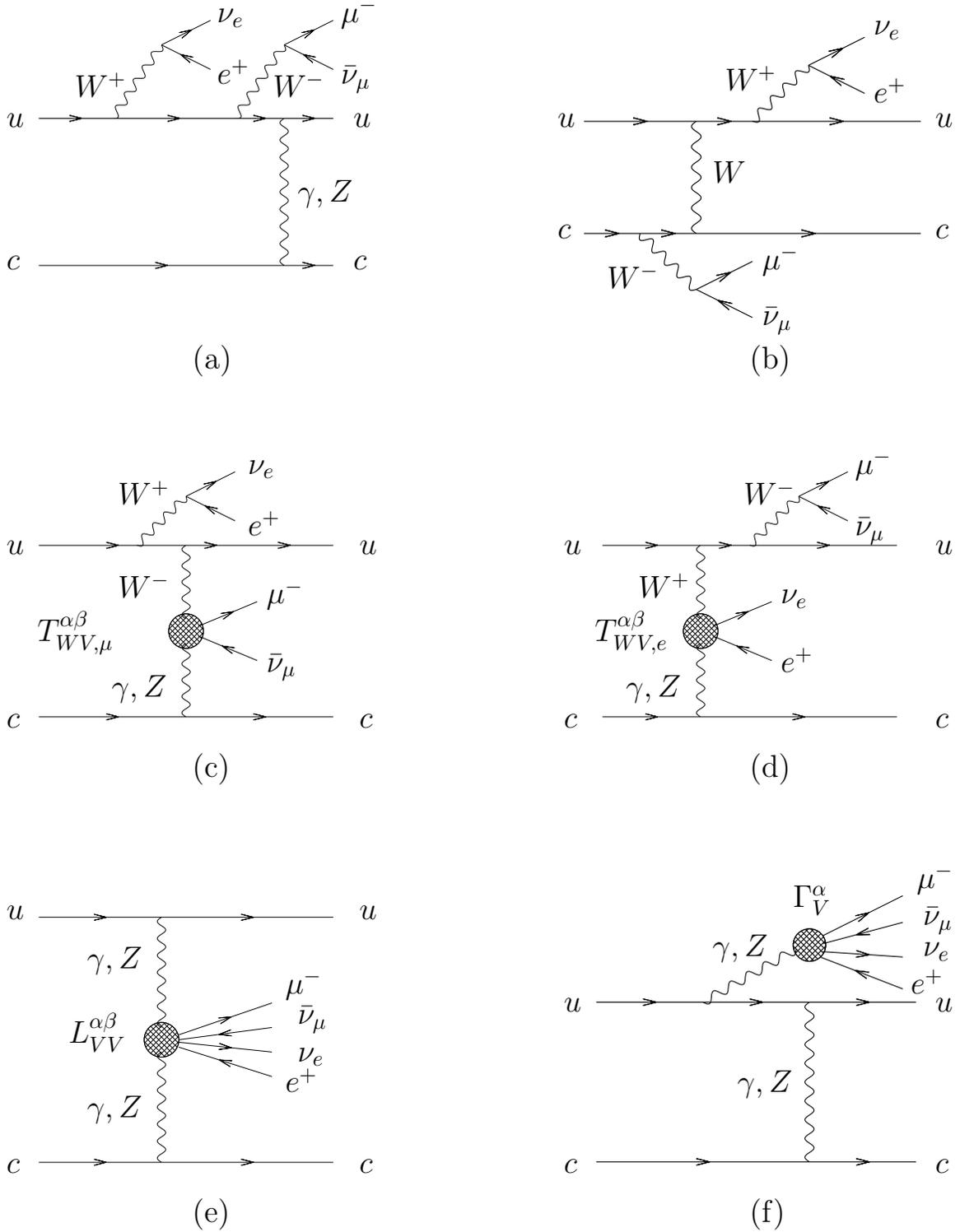}
\caption{\label{topologies} \emph{
Feynman-graph topologies contributing to the Born process 
$uc\rig uc e^+ \nu_e \mu^-\bar\nu_{\mu}$ in the $W^+W^- jj$ channel.}
}
\end{center}
\end{figure}

\begin{figure}[!ht]
\begin{center}
\includegraphics[angle=-90,width=0.9\textwidth]{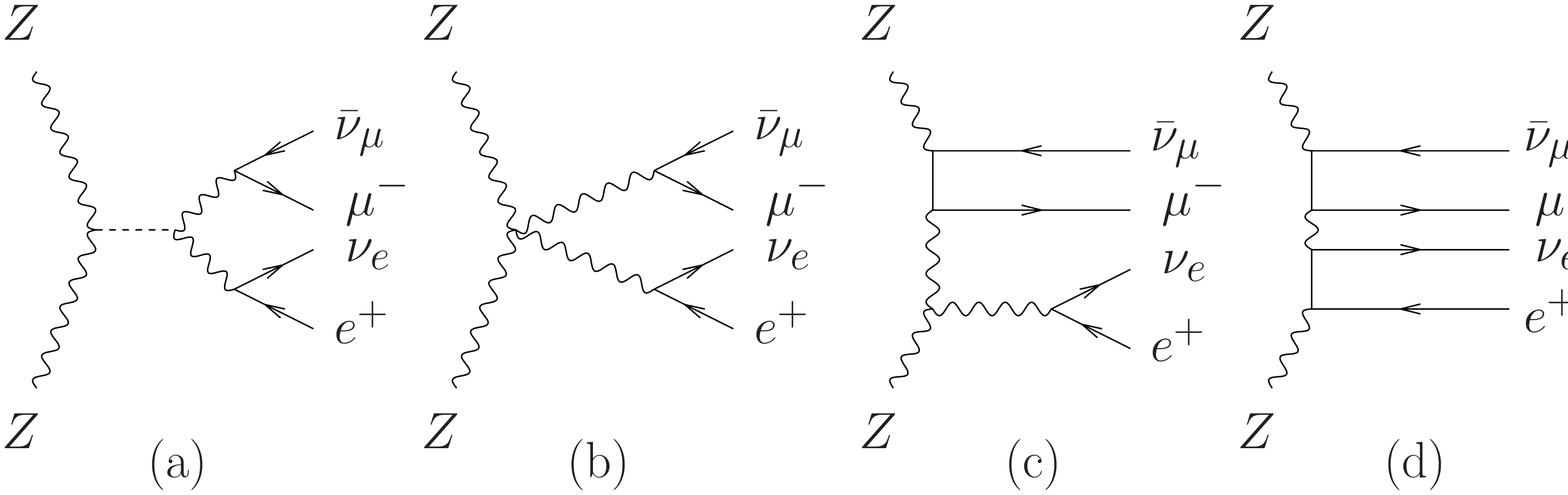}
\caption{\label{ltensor} \emph{
Representative diagrams contributing to the scattering amplitude $L_{ZZ}^{\alpha\beta}$ of Fig.~\ref{topologies}~(e) that describes the SM tree-level subprocess $ZZ\to e^+ \nu_e \mu^-\bar\nu_{\mu}$  with the $Z$ bosons carrying tensor indices $\alpha$ and~$\beta$.}
}
\end{center}
\end{figure}

The Feynman diagrams contributing to $pp\to \,e^+ \nu_e \mu^-\bar\nu_{\mu} jj$ can be grouped into six topologies, which are sketched in Fig.~\ref{topologies} for the $uc\to uc \,e^+ \nu_e \mu^-\bar\nu_{\mu}$ subprocess. 
%
%%%%%%%%%%%%%%%%%%%%%%%%%%%%%%%%%%%%%%%%%%%%%%%%%%%%%%%
%
%
%%%%%%%%%%%%%%%%%%%%%%%%%%%%%%%%%%%%%%%%%%%%%%%%%%%%%%%
%
The first two groups correspond to the emission of the external $W^+$ and $W^-$ bosons from the same (a) or different (b) quark lines. The remaining topologies are characterized by so-called ``leptonic tensors'' $T_{WV,\mu}^{\alpha\beta}$, $T_{WV,e}^{\alpha\beta}$, $L_{VV}^{\alpha\beta}$, and $\Gamma_{V}^{\alpha}$, which describe the tree-level amplitudes of the sub-processes 
$W^-V\to\,\mu^-\bar\nu_{\mu}$, $W^+V\to\,e^+ \nu_e$, $VV\to\,e^+ \nu_e \mu^-\bar\nu_{\mu}$, and $V\to\,e^+ \nu_e \mu^-\bar\nu_{\mu}$, respectively. In each case, $V$ stands for a virtual photon or $Z$~boson, and $\alpha$, $\beta$ are the tensor indices carried by the vector bosons. For each gauge boson with momentum $q$, mass $m$, and width $\Gamma$, the leptonic tensors include a propagator factor $1/(q^2-m^2+im\Gamma)$. 
The explicit structure of one of these leptonic tensors is depicted in Fig.~\ref{ltensor}, which shows some Feynman diagrams contributing to $L_{VV}^{\alpha\beta}$ within the SM. 
%
%%%%%%%%%%%%%%%%%%%%%%%%%%%%%%%%%%%%%%%%%%%%%%%%%%%%%%%
%
%
%%%%%%%%%%%%%%%%%%%%%%%%%%%%%%%%%%%%%%%%%%%%%%%%%%%%%%%
%

\begin{figure}[!b]
\begin{center}
\includegraphics[angle=-90,width=0.22\textwidth]{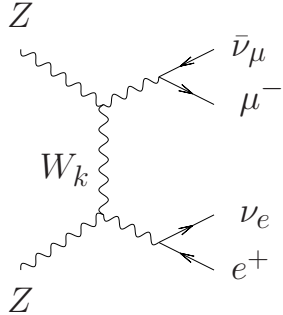}
\caption{\label{lktensor} \emph{
Diagrams contributing to the subprocess $ZZ\to e^+ \nu_e \mu^-\bar\nu_{\mu}$ in the context of the Warped Higgsless model.}
}
\end{center}
\end{figure}

Since the leptonic tensors parameterize all interactions in the gauge boson sector,  they can easily be generalized from the SM to a model including different gauge boson interactions. In particular, the Kaluza-Klein scenario we consider is consistently accounted for by adapting the gauge boson couplings entering $T_{WV,\mu}^{\alpha\beta}$, $T_{WV,e}^{\alpha\beta}$,  $L_{VV}^{\alpha\beta}$, $\Gamma_{V}^{\alpha}$, and complementing  the Feynman diagrams contributing to  $L_{VV}^{\alpha\beta}$.  The actual modification proceeds in two steps:
%
%%%%%%
%
\begin{enumerate}
\item All relevant couplings, masses and widths of the Kaluza-Klein model are calculated from the input parameters $m_{Z_1}$, $m_{W_1}$, and $R$  according to Eqs.~\gl{wtower}-\gl{3coup} in App.~\ref{app:whm}. Given the convergence of the sum rules, Eqs.~\gl{sumww1}-\gl{sumwz2}, the Kaluza-Klein spectrum is cut at  $k=n$. This procedure amounts to an explicit breaking of the higher-dimensional gauge invariance. In order to maintain a reasonable  high-energy behavior, the effect of neglecting the higher excitations is balanced by re-defining the quartic vertex coupling and the coupling of the $n$-th Kaluza-Klein mode to the SM bosons via the sum rules, Eqs.~\gl{sumww1}-\gl{sumwz2}. For the phenomenological studies of Secs.~\ref{sec:lhcpred} and \ref{sec:qcdcor},  states up to $W_6$ and $Z_{10}$ are included. The results have been found to be  stable with respect to changes in $n$. 
\item The leptonic tensor $L_{VV}^{\alpha\beta}$ for the $VV\to W^+W^-$ sub-amplitude of Fig.~\ref{topologies}~(e) is extended by Kaluza-Klein exchange contributions, while the Higgs contribution
of Fig.~\ref{ltensor} (a) is dropped. In Fig.~\ref{lktensor} the extra diagrams for the $ZZ\to  W^+W^-$ subprocess are depicted. 
\end{enumerate}
%
%%%%%%
%
%
\noindent Finite-width effects in massive vector boson propagators are
treated by means of a modified version of the complex mass scheme
throughout \cite{Denner:1999gp,Oleari:2003tc}: Vector-boson masses $m^2$
are globally replaced with $m^2-im\Gamma$, while a real value of
$\sin^2\theta_W$ is retained. 

For VBF $W^+W^-jj$ production, contributions from anti-quark initiated $t$-channel processes such as $\bar uc\to \bar uc \,e^+ \nu_e \mu^-\bar\nu_{\mu}$ or $\bar u\bar c\to \bar u\bar c \,e^+ \nu_e \mu^-\bar\nu_{\mu}$, which are obtained by crossing the quark-quark scattering diagrams shown above, are fully taken into account. In the same way, $u$-channel exchange diagrams are considered, which occur for diagrams obtained by interchange of identical initial- or final-state (anti-)quarks, such as in the $uu\to uu \,e^+ \nu_e \mu^-\bar\nu_{\mu}$ subprocess. However, interference effects of  $t$-channel with $u$-channel diagrams are neglected, as well as $s$-channel exchange diagrams which comprise the decay of a time-like vector boson into a pair of jets. In the phase-space regions where VBF can be observed experimentally, with widely separated jets of large invariant mass, the impact of the neglected contributions is very small \cite{Ciccolini:2007ec}. 

%%%%%%%%%%%%%%%%%%%%%%%%%%%%%%%%%%%%%%%%%%%%%%%%%%%%%%%
%
%
%%%%%%%%%%%%%%%%%%%%%%%%%%%%%%%%%%%%%%%%%%%%%%%%%%%%%%%
%
% 

While the amplitudes for VBF $W^+W^- jj$, $W^+Zjj$,  $W^-Zjj$, and  $ZZjj$ production have been implemented in \textsc{Vbfnlo} for specific input parameters of the Warped Higgsless model described above, the program may also be used in combination with externally calculated masses, couplings, and widths interfaced to  \textsc{Vbfnlo}. 

%
%--------------------------------------------------------------------
%			SUBSECTION
%--------------------------------------------------------------------
%
\subsection{Impact of the fermion sector}
For the model being consistent with measurements of the oblique
corrections it is necessary to let the fermions be spread out in the
bulk with a profile that minimizes their interaction with the
Kaluza-Klein towers \cite{Cacciapaglia:2004rb,Sekhar
  Chivukula:2005xm}. This allows for small contributions to the $S$
parameter and small couplings of the light fermions to the non-SM
KK gauge bosons. Existing bounds on the $T$ parameter further impose $R\lesssim 10^{-7}$GeV$^{-1}$. For VBF processes these contributions are at least suppressed by roughly \cite{Cacciapaglia:2004rb}
\be
{g_{f\bar{f}Z_2} g_{W_1W_1Z_2} \over g^2_{SM}}{m_Z^2\over m_{Z_2}^2}\lesssim 10^{-4}\,.
\ee
The error induced by neglecting the fermion interactions with the Kaluza-Klein gauge-boson sector in VBF should thus be negligible as compared to the uncertainties stemming from the dependence of the respective cross sections on factorization and renormalization scales and higher order corrections to be discussed below. 
We therefore disregard these fermion effects in the following. Consequently, all non-SM physics is encoded in the gauge sector. We calculate the widths of the  Kaluza-Klein excitations from their decays to the lower lying states, following \cite{Birkedal:2004au}. The considered model may thus be regarded as a type of the ``Higgsless-top Higgs scenario'' of Ref.~\cite{Cacciapaglia:2005pa}, which avoids inconsistencies in the heavy-quark sector \cite{Agashe:2006at,Cacciapaglia:2006gp}.

Entirely neglecting the fermion interactions with the Kaluza-Klein gauge bosons in principle leads to a residual growth of the VBF amplitude resulting from processes such as 
$W^+Z\rig e^+\nu_e$ or $Z,\gamma \rig e^+\nu_e\mu^-\bar\nu_\mu$ as a consequence of the modification of the $W_1W_1Z_1$ coupling from its 
SM value \cite{Baur:1987mt}. This amounts to explicitly violating gauge invariance when the first Kaluza-Klein modes, identified as the $W^\pm$ and $Z$ bosons of the SM, are coupled to fermions with SM strength in the leptonic tensors $T^{\alpha\beta}_{WV,\mu},
T^{\alpha\beta}_{WV,e}$, and $\Gamma^{\alpha}_V$ of Fig.~\ref{topologies}. In a fully consistent model these effects would be mended by respective contributions from a modified fermion sector, coupling to the Kaluza-Klein gauge bosons. 
%
%%%%%%%%%%%%%%%%%%%%%%%%%%%%%%%%%%%%%%%%%%%%%%%%%%%%%%%
%
\begin{figure}[!ht]
\begin{center}
\includegraphics[width=7.5cm,height=7.5cm]{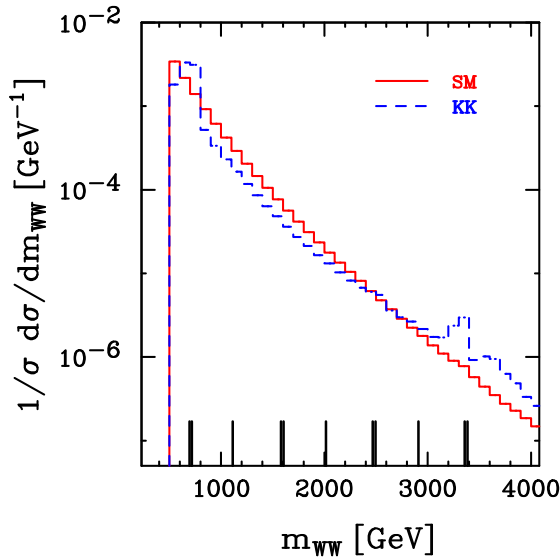}
\caption{\label{fig:invmasscomp} \emph{
Invariant mass distribution of the four leptons in  
$pp\rig e^+\nu_e\mu^-\bar{\nu}_\mu jj$ for the cut set-up of Sec.~\ref{sec:lhcpred} with an additional cut of $m_{WW}>500~\mathrm{GeV}$. The vertical solid lines mark the Kaluza-Klein $Z_k$ resonances in the $s$-channel of the gauge-boson scattering subprocess $VV \rig W^+W^-$. The uppermost Z resonance we have taken into account is enhanced as a consequence of balancing the sum rules \gl{sumww1}, \gl{sumww2}.
}}
\end{center}
\vskip -0.4cm
\end{figure}
%
%%%%%%%%%%%%%%%%%%%%%%%%%%%%%%%%%%%%%%%%%%%%%%%%%%%%%%%
%
%
%
In Fig.~\ref{fig:invmasscomp} we compare the high-energy behavior of a Kaluza-Klein model with $R=9.75\times~10^{-9}$  to the SM prediction for $W^+W^-jj$ production. With $m_H=120~\textnormal{GeV}$ and $m_{W^+W^-}\geq 130~\textnormal{GeV}$, the SM scenario is considered as prototype of a fully unitarized model without a visible Higgs resonance. Even at large invariant masses, no significant enhancement over the SM results occurs in the Kaluza-Klein distribution, which we take as an {\em a posteriori} justification of the method laid out above. The small change in the coupling induced by the Kaluza-Klein excitations leaves the overall gauge cancellations unaffected in practice. The visible excess over the SM estimate at high values of  $m_{W^+W^-}$ results from a residual logarithmic growth of the $VV\rig W^+W^-$ sub-amplitude at invariant masses large compared to the first Kaluza-Klein resonance \cite{Cacciapaglia:2004rb,Birkedal:2005yg}. Fig.~\ref{fig:invmasscomp} furthermore illustrates that a discovery of Kaluza-Klein $Z_k$ states with $k>3$ ($W^\pm_k$ with $k> 2$) is impossible in the considered VBF processes.
%
%--------------------------------------------------------------------
%			MAIN SECTION
%--------------------------------------------------------------------
%
\section{Vector boson scattering at the LHC}
\label{sec:lhcpred}
In this section we briefly discuss the phenomenological implications of the Warped Higgsless model described in Sec.~\ref{sec:hless}. We closely follow the discussion of Ref.~\cite{Zeppenfeld:2007ur}. For a more thorough study of the  discovery potential including a dedicated cut analysis, we refer to Ref.~\cite{Englert:2008tn}.

Throughout, we use the CTEQ6M parton distributions \cite{Pumplin:2002vw} with $\alpha_s(m_Z)=0.118$ at NLO, and the CTEQ6L1 set at LO. We choose $m_{Z_1}=91.188$GeV, $m_{W_1}=80.423~\textnormal{GeV}$ and $G_F=1.166\times 10^{-5}~\textnormal{GeV}^{-2}$ as electroweak input parameters and derive $\alpha_{\mathrm{QED}}$ and $\sin\theta_w$ via SM tree-level relations.
Jets are recombined from the final state partons via the $k_T$~algorithm \cite{Catani:1993hr} with resolution parameter $D=0.8$. We apply inclusive VBF cuts, requiring at least two jets of large transverse momentum, which are referred to as ``tagging jets'', with
\bee
\label{cuts1}
p_{T,j}^{\mathrm{tag}}\geq 20~\hbox{GeV}\,.
\eee
All jets need to lie in the rapidity range accessible to the detector, 
\bee
|\eta_j| \leq 4.5\,,
\eee
and are reconstructed from massless partons of pseudorapidity $|\eta| < 5$. The visible decay leptons are supposed to be hard and located at central rapidities,
\bee
p_{T,\ell}\geq 20~\hbox{GeV}\,, \qquad |
\eta_{\ell}|\leq 2.5\,. 
\eee 
They are furthermore required to be well-separated from each other and from the jets,
\bee
\Delta R_{\ell\ell}\geq 0.2\,, \qquad 
\Delta R_{j\ell}\geq 0.4\,,
\eee
where $\Delta R_{\ell\ell}$ is the lepton-lepton and $\Delta R_{j\ell}$ the jet-lepton separation in the azimuthal angle-pseudorapidity-plane. Via the $\Delta R_{\ell\ell}$ cut, collinear singularities from the virtual photon decay $\gamma\rig \ell^+\ell^-$ are avoided. 
%
%%%%%%%%%%%%%%%%%%%%%%%%%%%%%%%%%%%%%%%%%%%%%%%%%%%%%%%
%
\begin{table}[t!]
\begin{center}
\begin{tabular}{ | c | c | c | c | c | }
\hline
scenario & $R$~[GeV$^{-1}$] & $m_{W_2}$~[GeV]& $m_{Z_2}$~[GeV]& $m_{Z_3}$~[GeV] \\
\hline
\hline
A & $9.75\times 10^{-9}$ & 700 & 695 & 718\\
%\hline
B & $10^{-19}$ & 1190 & 1187 & 1200 \\
\hline
\end{tabular}
\caption{\label{tab:kk}\emph{
Parameters of the Warped Higgsless Kaluza-Klein scenarios used in the simulation.}}
\end{center}
\end{table}
%
%
%%%%%%%%%%%%%%%%%%%%%%%%%%%%%%%%%%%%%%%%%%%%%%%%%%%%%%%
%
\begin{table}[t!]
\begin{center}
\begin{tabular}{ | c | c | c | c | }
\hline
process & SM & KK (A) & KK (B) \\
\hline
\hline
$W^+W^-\rig e^+\nu_e\mu^-\bar{\nu}_{\mu}$ & 1.695 & 2.28 & 2.03\\
%\hline
$W^+Z \rig e^+\nu_e\mu^+\mu^-$ 		  & 0.184 & 0.35 & 0.24 \\
$W^-Z\rig e^-\bar{\nu}_e\mu^+\mu^-$ 	  & 0.102   & 0.19 & 0.13\\
%\hline
$ZZ\rig e^+e^-\nu_{\mu}\bar{\nu}_{\mu}$   & 0.132 & 0.17& 0.16 \\
$ZZ\rig e^+e^-\mu^+\mu^-$ 		  & 0.04 & 0.06& 0.06 
\\
\hline
\end{tabular}
\caption{\label{crossecslo}\emph{
Cross sections (in fb) for various $VVjj$ production processes in the SM and the Warped  Higgsless Kaluza-Klein scenarios of Tab.~\ref{tab:kk} within the cuts of Eqs.~\gl{cuts1}-\gl{cutsf}. For the SM predictions, additionally Eq.~\gl{eq:mvv} has been imposed. Statistical errors are below $0.1\%$.
}}
\end{center}
\vskip-0.4cm
\end{table}
%
%%%%%%%%%%%%%%%%%%%%%%%%%%%%%%%%%%%%%%%%%%%%%%%%%%%%%%%
%
%
We impose a large rapidity gap between the tagging jets and demand that they be  detected in opposite detector hemispheres,  
\bee
\Delta\eta_{jj}=|\eta^{\mathrm{tag}}_{j_1}-\eta^{\mathrm{tag}}_{j_2}|>4\,,\quad 
\eta^{\mathrm{tag}}_{j_1}\times \eta^{\mathrm{tag}}_{j_2}<0\,,
\eee
with a large invariant mass, 
\bee
m^{\mathrm{tag}}_{j_1j_2}\geq 600~\hbox{GeV}\,.
\eee
In addition, the leptons are required to fall into the rapidity gap between the two tagging jets, 
\bee
\label{cutsf}
\min\{\eta^{\mathrm{tag}}_{j_1},\eta^{\mathrm{tag}}_{j_2}\}\leq \eta_l\leq \max\{\eta^{\mathrm{tag}}_{j_1},\eta^{\mathrm{tag}}_{j_2}\}\,.
\eee
If not specified otherwise, factorization and renormalization scales, $\mu_R$ and $\mu_F$, for the upper and lower fermion lines are set equal to the momentum transfer $Q$ carried by the vector boson attached to the respective quark line in VBF graphs as in Fig.~\ref{topologies}~(e) \cite{Zeppenfeld:2007ur}.   
At tree-level, we include detector-resolution effects based on Gaussian smearing of the events according to \cite{Pi:2006ct,Heister:2006cu} throughout. 

As representative Warped Higgsless models, 
we consider the two Kaluza-Klein spectra sketched in Tab.~\ref{tab:kk}.
Scenario~A corresponds to a relatively light Kaluza-Klein spectrum, which postpones unitarity violations in gauge boson scattering reactions to an energy range of several TeV. In scenario~B, the location of the Planck brane is identified with the fundamental scale of the RSI model, $R=10^{-19}$~GeV$^{-1}\sim 1/M^{d=4}_{\mathrm{Pl}}\sim 1/ M^{d=5,\mr{RSI}}_{\mathrm{Pl}}$, with a partial wave unitarity violation
scale of about $2.8~\mathrm{TeV}$.
For reference, we also compute SM predictions, setting $m_H=120$~GeV and 
\bee
\label{eq:mvv}
m_{VV}>130~\mr{GeV}\,.
\eee

Leading order (LO) cross sections for various SM and Kaluza-Klein $VVjj$ production processes are listed in Tab.~\ref{crossecslo} within the selection cuts of Eqs.~\gl{cuts1}-\gl{cutsf}. For the SM predictions, we additionally imposed Eq.~\gl{eq:mvv}.
The process-specific features of the individual VBF channels are discussed in the following. 
%
%--------------------------------------------------------------------
%			SUBSECTION
%--------------------------------------------------------------------
%
\subsection{${W}^+W^-jj$ production}
\label{wpwmsec}
In the energy range accessible at the LHC, the $W^+W^-jj$ channel is sensitive to the two Kaluza-Klein resonances $Z_2$ and $Z_3$. Irrespective of the details of the Kaluza-Klein spectrum, production cross sections are considerably enhanced with respect to the SM, cf.\ Tab.~\ref{crossecslo}. 
%
%
%
%%%%%%%%%%%%%%%%%%%%%%%%%%%%%%%%%%%%%%%%%%%%%%%%%%%%%%%
%
\begin{figure}[!t]
\begin{center}
\includegraphics[width=7.5cm,height=7.5cm]{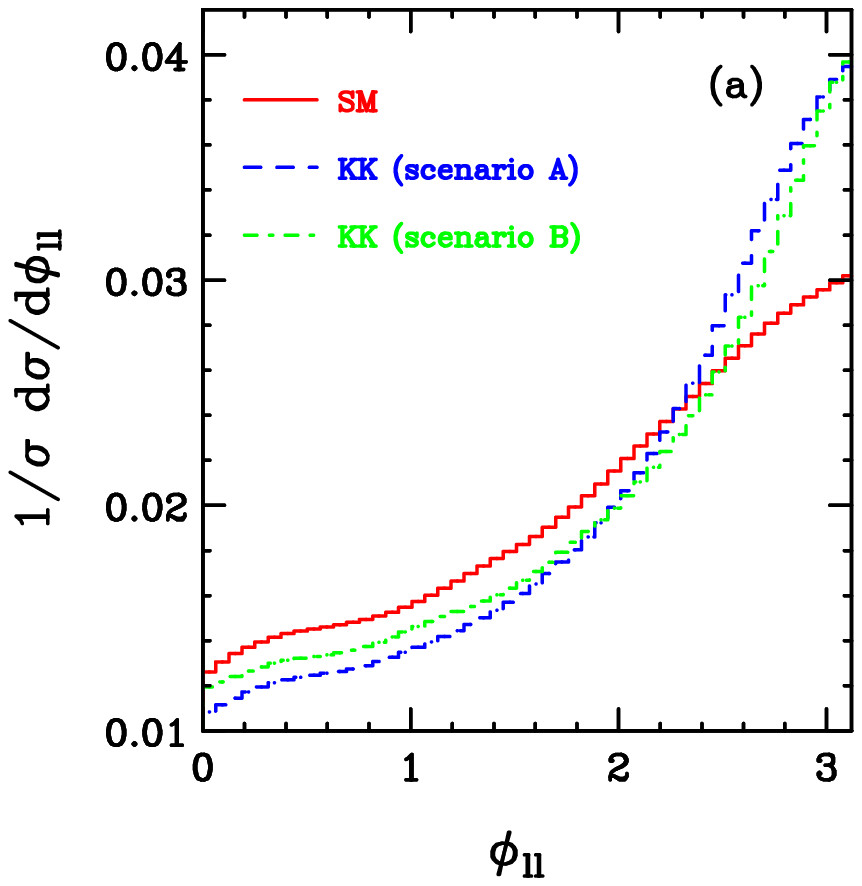}
\hfill
\includegraphics[width=7.5cm,height=7.5cm]{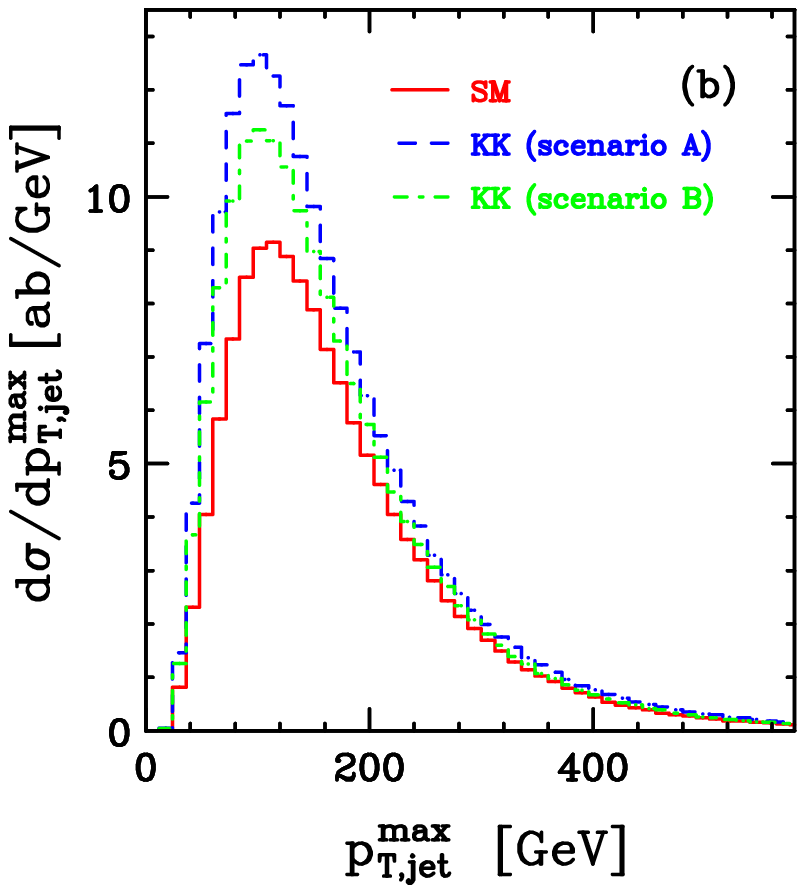}
\caption{\label{wwplot}\emph{Azimuthal angle distribution of the two decay leptons (a) and transverse momentum distribution of the hardest jet (b) for  $pp\rig W^+W^-jj$. Shown are predictions for the SM (red, solid), and for the two Higgsless scenarios A~(blue, dashed) and B~(green, dot-dashed).}}
\end{center}
\vskip -0.4cm 
\end{figure}
%
%%%%%%%%%%%%%%%%%%%%%%%%%%%%%%%%%%%%%%%%%%%%%%%%%%%%%%%
%
%
In scenario~A, the wave functions of the $Z_2$ and the $Z_3$  excitation are very similar. Thus, they contribute almost equal shares to the sum rules of Eqs.~\gl{sumww1}-\gl{sumwz2}, and have approximately equal couplings to the $W^\pm_1$ bosons.  
For the heavier Kaluza-Klein spectrum of scenario~B, the couplings are controlled by the sum rule of Eq.~\gl{sumww2}, so that $Z_3$ couples to the SM modes more weakly than the first excitation. In particular, the coupling $g_{W_2W_1Z_1}$ of scenario~B amounts to only $57\%$ of $g_{W_2W_1Z_1}$ in scenario~A. Together with the decreased phasespace available, this leads to smaller cross sections for the heavier Kaluza-Klein mass spectrum than for the light scenario.
A heavy particle in the $s$-channel of the leptonic tensor $L^{\alpha\beta}$ affects the angular correlations of the visible leptons such as the azimuthal angle between the two charged leptons shown in Fig.~\ref{wwplot}~(a). 
The leptons tend to be back-to-back as they undergo large boosts resulting from the large momenta of the $W^\pm$ bosons in the Kaluza-Klein $Z_k$ rest frame. Accordingly, the leptons are somewhat harder than in the SM case, which is balanced by the slightly softer jet-$p_T$ spectrum, cf.\ Fig.~\ref{wwplot}~(b). 
%
%--------------------------------------------------------------------
%			SUBSECTION
%--------------------------------------------------------------------
%
\subsection{\label{zzsec}$Z Z jj$ production}
In the absence of a Higgs boson in the particle spectrum, $ZZ\rig ZZ$ scattering does not occur at tree-level. Consequently, the $ZZjj$ channels are rather insensitive to the choice of $R$ and the
corresponding Kaluza-Klein mass spectrum, as $W_k$ excitations enter only via $t$-channel exchange contributions in $W^+W^-\to ZZ$ sub-amplitudes. Indeed, the LO $ZZjj$ production cross sections listed in Tab.~\ref{crossecslo} are comparable for the two scenarios of Tab.~\ref{tab:kk}. 
Nonetheless, larger cross sections are obtained in Higgsless models than in the
SM due to the lack of scalar Higgs boson exchange diagrams. In the SM, such
contributions to $W^+W^- \rig ZZ$ scattering differ in sign from pure gauge boson exchange terms and thus cause a decrease of the full scattering amplitude.   

\begin{figure}[t!]
\begin{center}
\includegraphics[width=7.5cm,height=7.5cm]{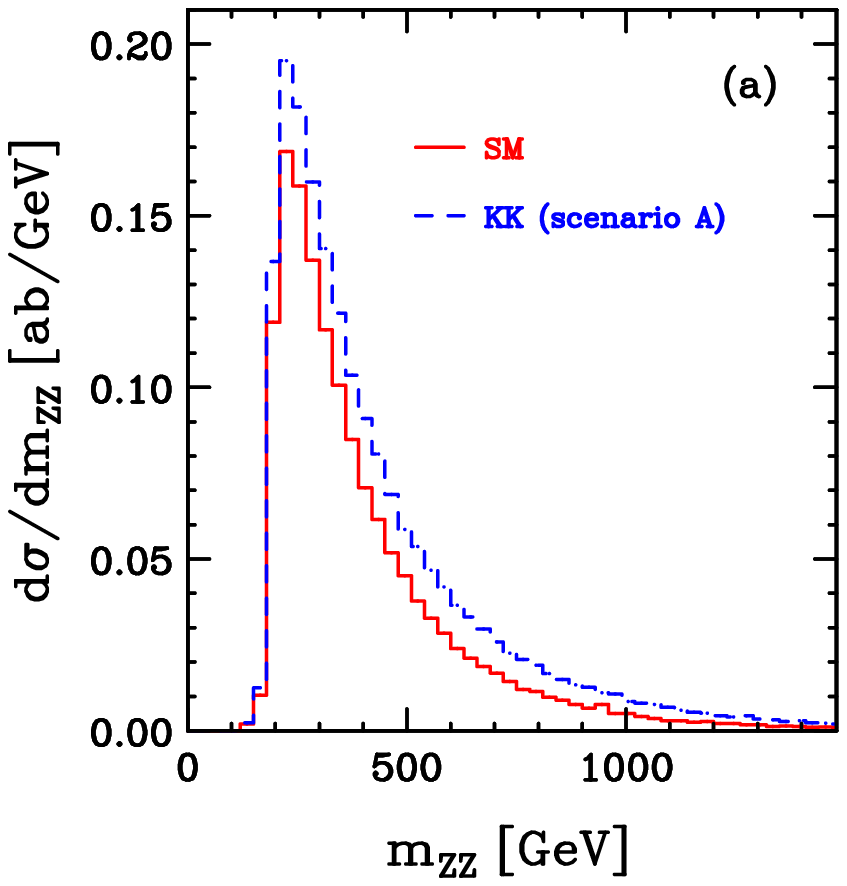}
\hfill
\includegraphics[width=7.5cm,height=7.5cm]{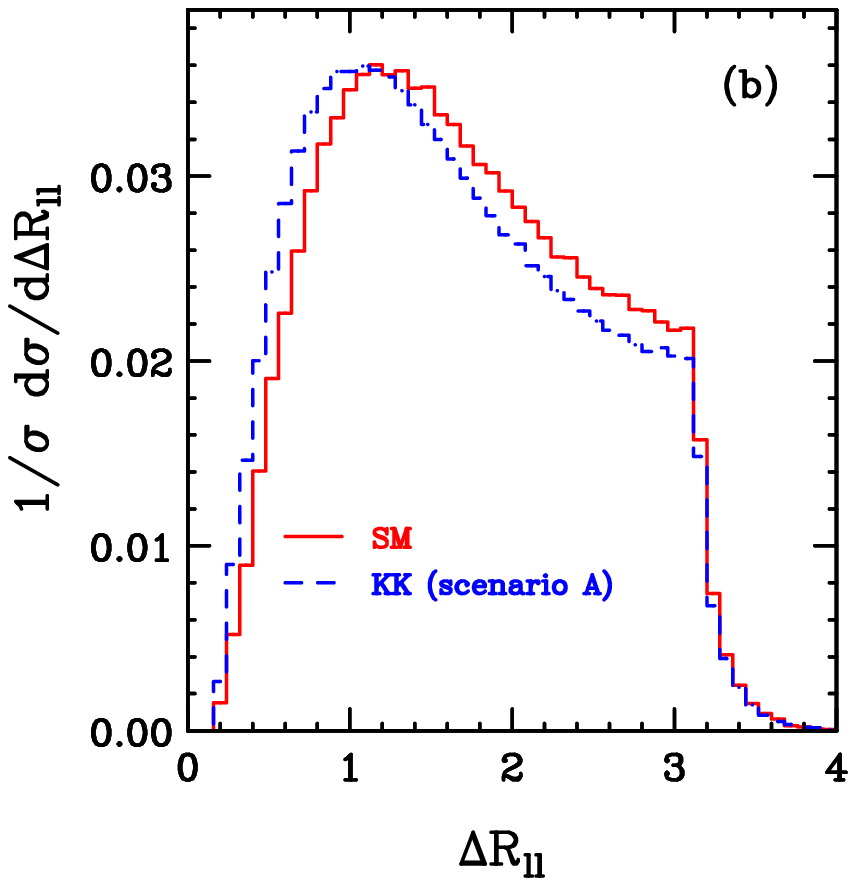}
\caption{\label{zzplot}\emph{
Invariant mass distribution of the four decay leptons in $pp\rig ZZjj\rig 4\ell jj$ (a)  
and $\Delta R_{\ell\ell}$ separation of the two charged decay leptons in $pp\rig ZZjj\rig
2\ell \slashed{p}_Tjj$ (b). Shown are predictions for the SM (red, solid) and for the  Warped Higgsless scenario A of Tab.~\ref{tab:kk}~(blue, dashed).}
}
\end{center}
\vskip -0.4cm 
\end{figure}

In Fig.~\ref{zzplot}~(a) 
%
%
%%%%%%%%%%%%%%%%%%%%%%%%%%%%%%%%%%%%%%%%%%%%%%%%%%%%%%%
%
%
%%%%%%%%%%%%%%%%%%%%%%%%%%%%%%%%%%%%%%%%%%%%%%%%%%%%%%%
%
%
this feature is illustrated by the invariant mass distribution of the decay
leptons in the $pp \rig Z Zjj\rig 4\ell jj$ channel, which is slightly larger in the Warped Higgsless scenario than in the SM, but very similar in shape. In contrast, the lego-plot separation of the two charged leptons in  $pp\rig ZZjj\rig
2\ell\slashed{p}_Tjj$, shown in Fig.~\ref{zzplot}~(b), changes shape as well, tending to lower values in the Kaluza-Klein model than in the SM. 
%
%
%%%%%%%%%%%%%%%%%%%%%%%%%%%%%%%%%%%%%%%%%%%%%%%%%%%%%%%
%
\begin{figure}[t!]
\begin{center}
\includegraphics[width=7.5cm,height=7.5cm]{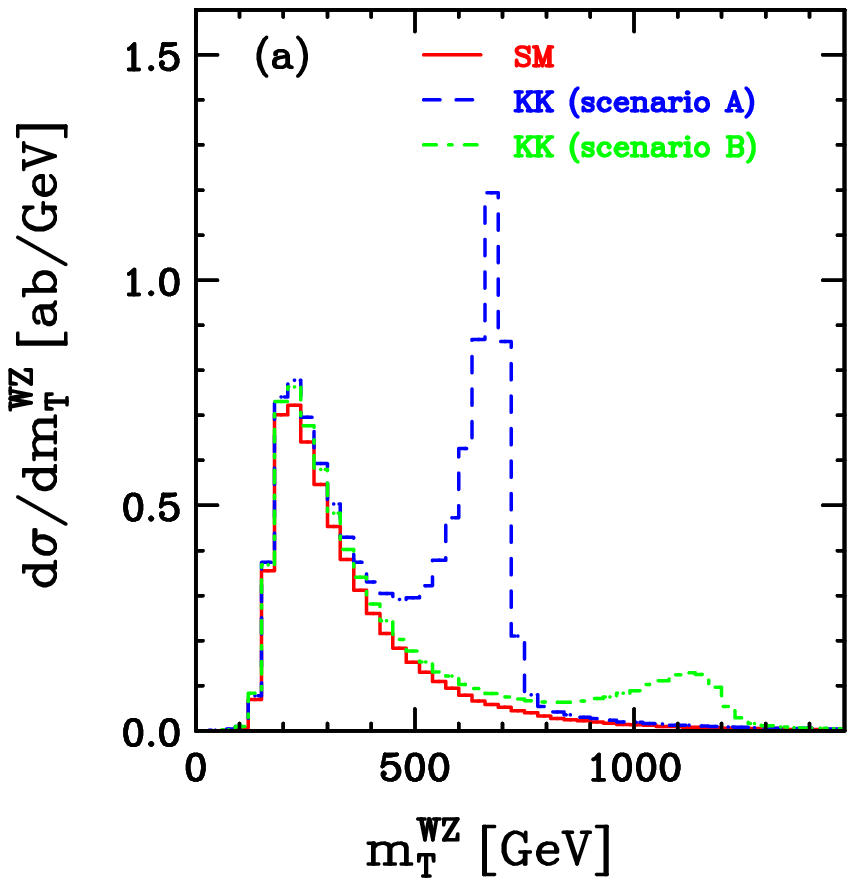}
\hfill
\includegraphics[width=7.5cm,height=7.5cm]{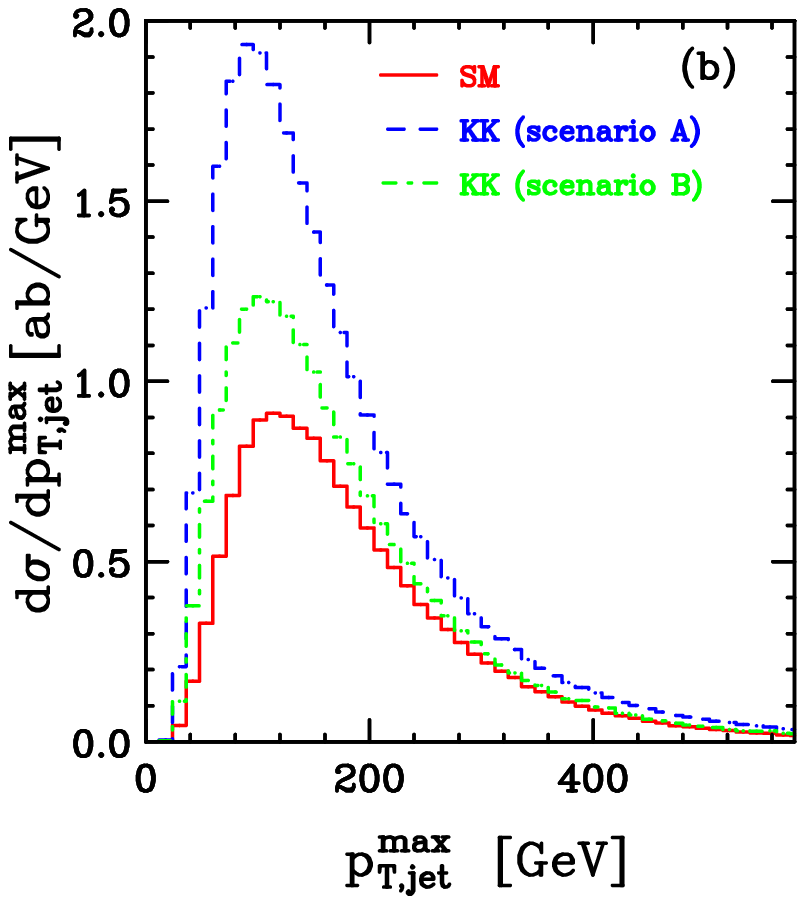}
\caption{\label{wzplot}\emph{
Transverse cluster mass distribution (a) and transverse momentum distribution of the hardest tagging jet (b) for $pp\rig W^+Zjj$. Shown are predictions for the SM (red, solid), and for the two Higgsless scenarios A~(blue, dashed) and B~(green, dot-dashed).}}
\vskip 3cm
\includegraphics[width=7.5cm,height=7.5cm]{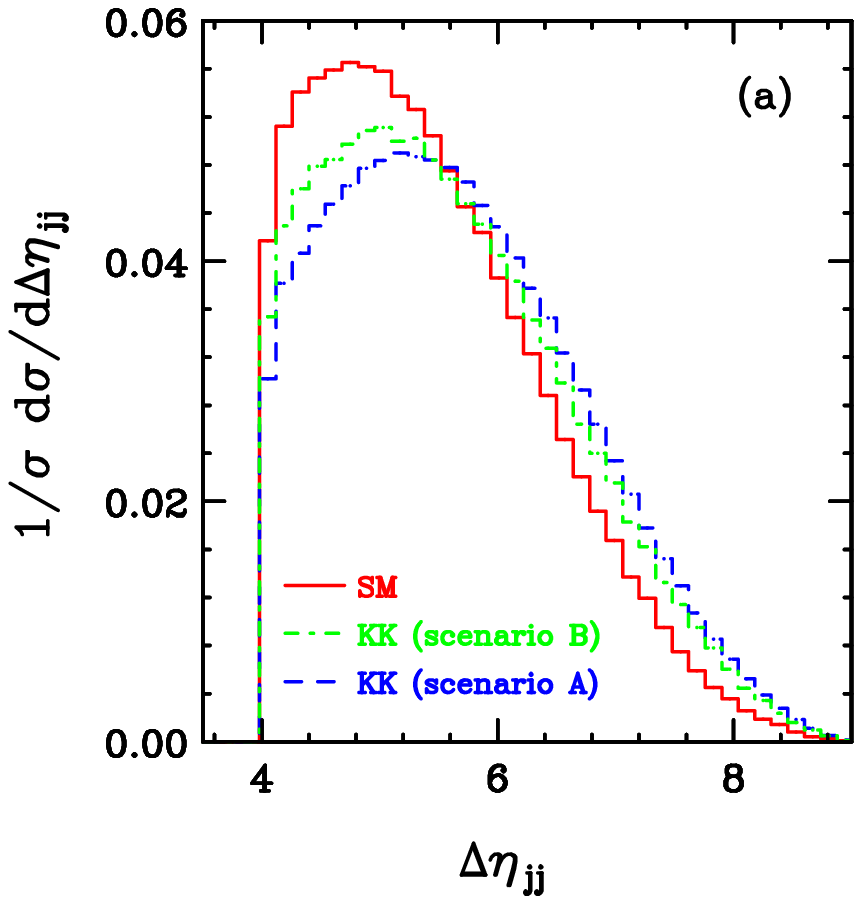}
\hfill
\includegraphics[width=7.5cm,height=7.5cm]{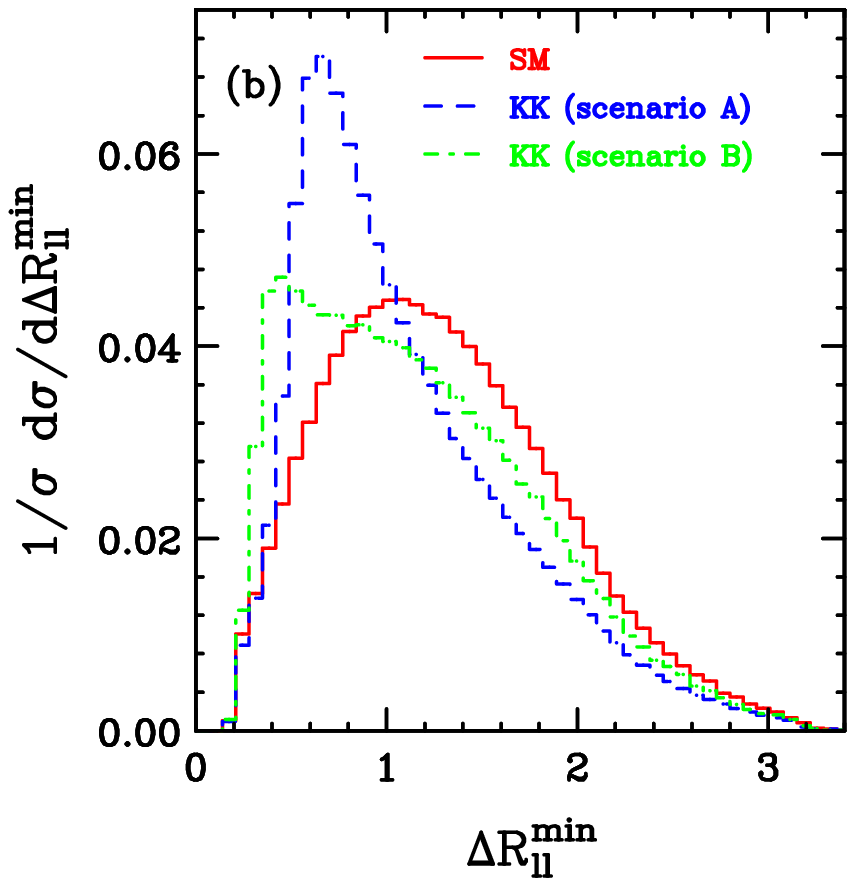}
\caption{\label{wzplot2}\emph{
Rapidity-difference distribution of the tagging jets (a) and minimal $\Delta R_{\ell\ell}$ separation of the charged leptons (b) for $pp\rig W^+Zjj$. Shown are predictions for the SM (red, solid), and for the two Higgsless scenarios A~(blue, dashed) and B~(green, dot-dashed).}}
\end{center}
\vskip -0.4cm
\end{figure}
%
%%%%%%%%%%%%%%%%%%%%%%%%%%%%%%%%%%%%%%%%%%%%%%%%%%%%%%%
%
%
%--------------------------------------------------------------------
%			SUBSECTION
%--------------------------------------------------------------------
%
\subsection{\label{wzchann}${W}^\pm Zjj$ production}
For distinguishing the Warped Higgsless scenario from other models of electroweak symmetry breaking, the $W^\pm Zjj$ channels with their striking resonance structure are most appropriate. 
In the following, we focus on the $W^+Zjj$ mode. The $W^-Zjj$ channel exhibits distributions of very similar shape \cite{Englert:2008tn}, but an approximately 50\% smaller cross section (cf.~Tab.~\ref{crossecslo}), due to the size of the parton distribution functions entering the dominant subprocess cross sections.

Because of the invisible neutrino in the final state, the invariant mass of the $W^\pm Z$ system cannot be fully reconstructed. However, the mass of the Kaluza-Klein resonance can be deduced from the Jacobian peak in the transverse cluster mass,  defined by \cite{Bagger:1993zf}
\bee
\label{transvwp}
m^2_T(WZ)= \left(\sqrt{m^2(\ell\ell\ell)+\vec{p}_{T}(\ell\ell\ell)^{\,2}}+ |\slashed{p}_T|\right)^2
 - \left(\vec{p}_{T}(\ell\ell\ell)+\slashed{\vec{p}}_T\right)^2\,.
\eee
We depict $m_T(WZ)$  in Fig.~\ref{wzplot}~(a) for the SM and the two Kaluza-Klein mass spectra of Tab.~\ref{tab:kk}. For both parameter choices, the $W_2^+$ excitation manifests itself in a pronounced resonance peak, which clearly distinguishes these scenarios from the SM. 

In contrast to the previously discussed $W^+W^-jj$ and $ZZjj$ channels, 
in the $W^\pm Zjj$ mode Kaluza-Klein resonances cause sizeable modifications of the jet distributions. 
The transverse momentum distribution of the hardest tagging jet, shown in Fig.~\ref{wzplot}~(b), peaks at smaller values of $p_{T,j}^\mr{tag}$ in the Higgsless scenarios than in the SM. 
At the same time, the rapidity differences of the tagging jets, $|\eta^{\mathrm{tag}}_{j_1}-\eta^{\mathrm{tag}}_{j_2}|$, depicted in Fig.~\ref{wzplot2}~(a), and of the tagging jets and the charged leptons, 
$|\eta^{\mathrm{tag}}_{j}-\eta_{\ell}|$, are shifted to larger values. This is due to the leptons being produced slightly more centrally because of the small boost of the $W_2$ resonance. The charged leptons, on the other hand, tend to be closer to each other, resulting in smaller values of $\Delta R_{\ell\ell}$, 
cf.\ Fig.~\ref{wzplot2}~(b).
%
%--------------------------------------------------------------------
%			MAIN SECTION
%--------------------------------------------------------------------
%
\section{Impact of NLO-QCD corrections}
\label{sec:qcdcor}
Including the unbroken QCD gauge group $\mr{SU}(3)_C$ in the bulk of the Warped Higgsless model yields testable predictions \cite{Davoudiasl:2003me} beyond the electroweak sector: As a consequence of the unbroken gauge symmetry in the four-dimensional effective theory, gluonic excitations in the TeV~range emerge. 
However, in the kinematic regime of gauge boson scattering at the LHC, the impact of the first massive gluonic excitation is negligible.  
It is thus reasonable to disregard contributions from massive Kaluza-Klein gluons when contemplating QCD corrections to weak boson scattering processes in the context of Higgsless models.
Within this approximation, NLO-QCD corrections to VBF in Higgsless models can be determined in complete analogy to the SM \cite{Zeppenfeld:2007ur}. Doing so, we treat QCD as entirely decoupled from the strongly interacting sector that breaks the electroweak symmetry. 

\begin{figure}
\begin{center}
\includegraphics[width=15cm]{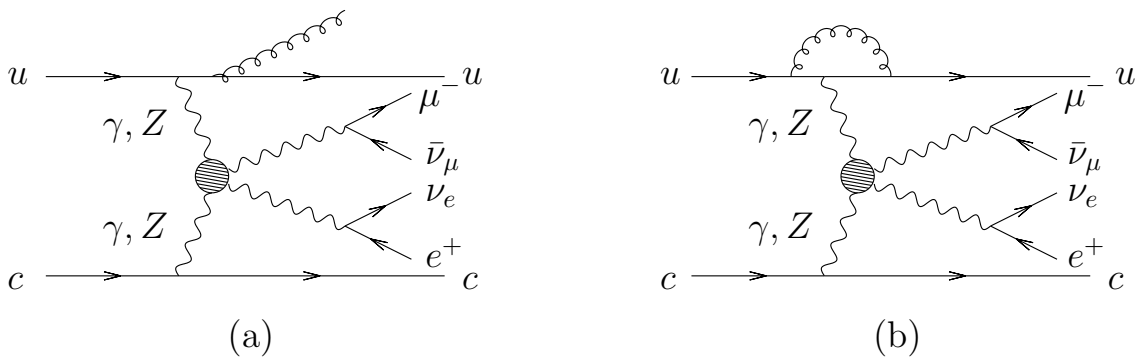}
\caption{\label{feyn-nlo}\emph{
Representative diagrams contributing to the real emission (a) and the virtual corrections (b) to $uc\to uc\,e^+\nu_e \mu^-\bar\nu_{\mu}$. The shaded area contains different intermediate states.
}}
\end{center}
\end{figure}

VBF processes are characterized by a particularly simple QCD structure due to the color-singlet nature of the gauge bosons exchanged in the $t$-channel. This feature is not spoiled by the inclusion of Kaluza-Klein excitations in the leptonic tensors of the gauge-boson scattering sub-amplitudes. The form of the NLO-QCD corrections to VBF in Warped Higgsless models is thus identical to the SM and can readily be adapted from Ref.~\cite{Zeppenfeld:2007ur}.
We employ the dipole subtraction formalism \cite{Catani:1996vz}, which requires the  computation of real emission and virtual corrections to the Born amplitude $\mathcal{M}_B$, and of counter terms to absorb singularities which emerge in intermediate steps of the calculation. The calculation is performed in the dimensional reduction scheme \cite{DR_citation} in $d=4-2\epsilon$ dimensions.  
The real emission contributions are obtained by attaching an extra gluon to the (anti-)quark initiated Born processes in all possible ways (e.g.\ $uc\to ucg\,e^+\nu_e \mu^-\bar\nu_{\mu}$, see Fig.~\ref{feyn-nlo}~(a)) and furthermore including channels with a gluon in the initial state, such as $gc\to u\bar uc\,e^+\nu_e \mu^-\bar\nu_{\mu}$. 
For the virtual contributions, triangle, box, and pentagon corrections to either the upper or the lower quark line have to be considered. An exemplary class of diagrams has been depicted in Fig.~\ref{feyn-nlo}~(b). 
Graphs with a gluon attached to both the upper and the lower quark line vanish at order $\alpha_s$ since they do not interfere with the color-singlet Born amplitude, within our approximations. Upon summing all virtual corrections, we find
\bee
\label{eq:virtual_born}
2 \,\mr{Re}\, [{\cal M}_V{\cal M}_B^* ]
&=& |{\cal M}_B|^2\, \frac{\alpha_s(\mu_R)}{2\pi}\, C_F
\left(\frac{4\pi\mu_R^2}{Q^2}\right)^\epsilon \Gamma(1+\epsilon)\nonumber \\
 &&\times
\left[-\frac{2}{\epsilon^2}-\frac{3}{\epsilon}+\frac{\pi^2}{3}-7\right]
+2 \,\mr{Re}\, [ \widetilde{\cal M}_V{\cal M}_B^* ] \,,
\eee
where $C_F=4/3$, and $\widetilde{\cal M}_V$ is a completely finite remainder. 
The poles in the virtual contribution are canceled by respective singularities in the phase-space integrated counter terms. For the evaluation of the five-point tensor integrals entering $\widetilde{\cal M}_V$, we resort to the procedure of Ref.~\cite{Denner:2002ii}. Two-, three- and four-point tensor integrals are evaluated by a standard Passarino-Veltman reduction \cite{Passarino:1978jh}. 

With the NLO-QCD corrections being of exactly the same form as in the SM, 
the implementation of the Warped Higgsless model into the framework of \textsc{Vbfnlo} is rather straightforward: the same leptonic tensors can be used for the $\mathcal{O}(\alpha_s)$ corrections
and the LO calculation and these leptonic tensors fully contain the BSM effects.
 \textsc{Vbfnlo} allows for the computation of cross sections and arbitrary infrared-safe distributions at order $\mathcal{O}(\alpha^6\alpha_s)$ accuracy within experimentally feasible selection cuts in the same manner as the LO version of the program. 

Since the signatures of the Higgsless models we consider are most distinctive in the $W^\pm Z jj$ channel, we study NLO-QCD corrections for this production mode within the settings of Sec.~\ref{sec:lhcpred} as an example. The impact of the NLO contributions in the Kaluza-Klein scenario turns out to be comparable to the SM \cite{Zeppenfeld:2007ur}.  
To quantify the size of the NLO-QCD corrections, we consider 
the differential $K$~factor
\bee
\label{diffk}
K(\mathcal{O}) = \ta{\sigma^{\mathrm{NLO}}}{\mathcal{O}} \bigg/  \ta{\sigma^{\mathrm{LO}}}{\mathcal{O}} \,.
\eee
In \cite{Zeppenfeld:2007ur} it was pointed out that in the SM a suitable choice of the factorization scale $\mu_F$ can help in obtaining LO shapes which closely resemble the NLO predictions for VBF processes. In particular, choosing $\mu_F=Q$ was found to result in LO distributions very similar to the NLO predictions and yield $K$~factors close to one. NLO results, on the other hand, are barely sensitive to the scale choice. 
This feature remains unaffected by the inclusion of new interactions in the color-neutral gauge boson sector as in Warped Higgsless models, and is thus present also in the scenario we consider, independently of the actual mass spectrum of the underlying Kaluza-Klein tower. 

\begin{figure}
\begin{center}
\includegraphics[width=7.5cm,height=7.5cm]{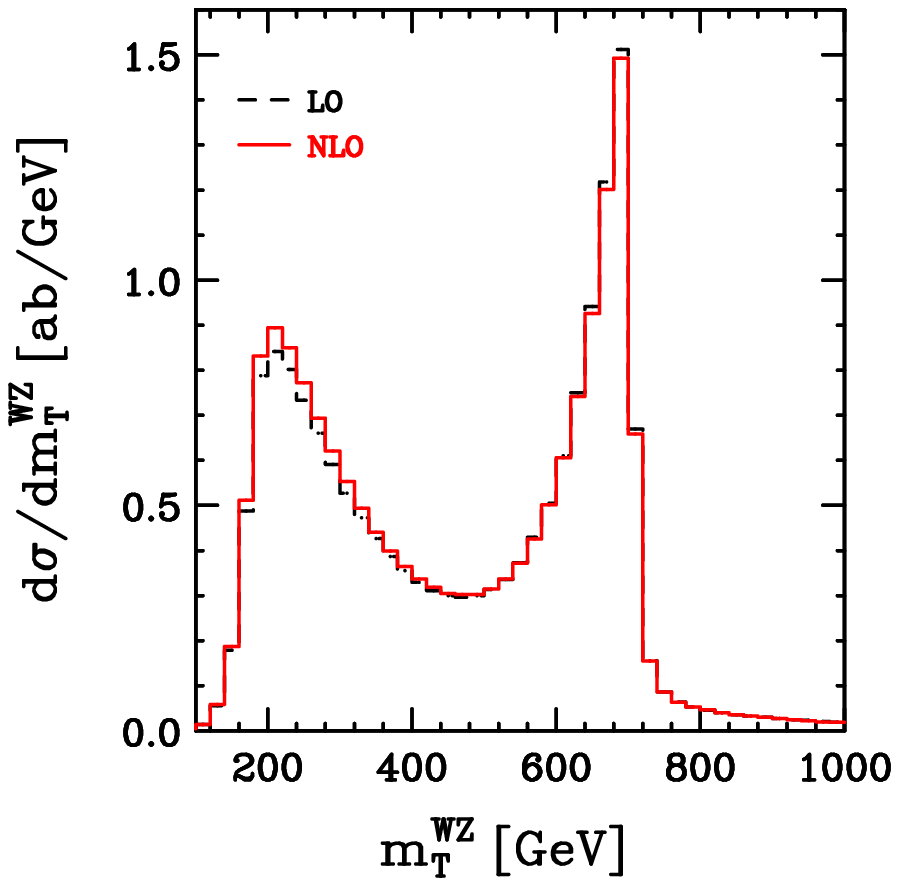}
\hfill
\includegraphics[width=7.5cm,height=7.5cm]{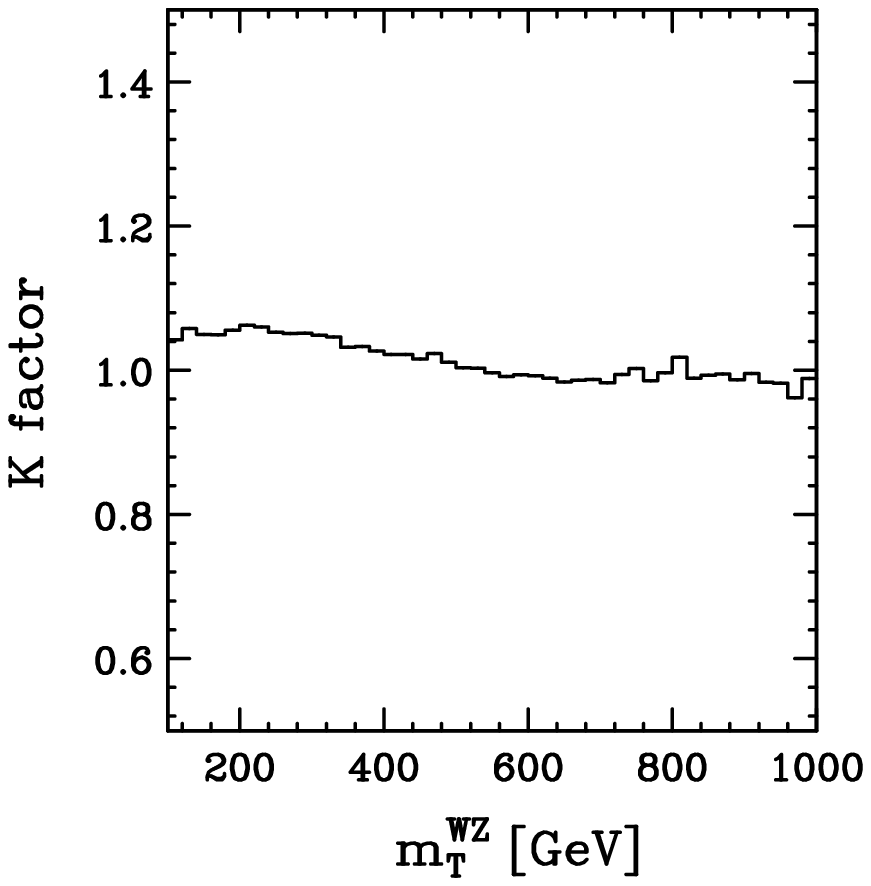}
\caption{\label{mtnlo}\emph{
LO (black dashed line) and NLO (red solid line) distribution of the transverse cluster mass of the  $W^+Z$ system in a Warped Higgsless scenario A of Tab.~\ref{tab:kk}  and differential $K$~factor. Scales are set to 
$\mu_R=\mu_F=Q$.}}
\end{center}
\end{figure}

To better illustrate the scale dependence of the LO and NLO results, Tab.~\ref{tab:scales} gives cross sections and $K$~factors for the Kaluza-Klein scenario A of Tab.~\ref{tab:kk} within the cuts of Eqs.~\gl{cuts1}--\gl{cutsf} for different choices of the factorization scale. The renormalization scale $\mu_R$, which enters only at NLO, is taken to be equal to the factorization scale ($\mu=\mu_F=\mu_R$). 
\begin{table}
\begin{center}
\begin{tabular}{|c | c|c|c|}
\hline
Scale $\mu$ &
$\sigma^\mr{LO}$~[fb]   &
$\sigma^\mr{NLO}$~[fb] &
$K$ factor
\\
\hline
\hline
 $(m_W+m_Z)/2$  & 0.359  & 0.355 &  0.989\\
 $Q$ 		& 0.349  & 0.356 &  1.020\\ 
 $m_{W_2}$ 	& 0.283  & 0.346 &  1.223 \\
\hline
\end{tabular}
\caption{\emph{
\label{tab:scales}
Cross sections and $K$~factors for $W^+Zjj$ production in the Warped Higgsless scenario A of Tab.~\ref{tab:kk} within the cuts of Eqs.~\gl{cuts1}-\gl{cutsf} for different choices for the factorization and renormalization scales. The statistical errors are below 0.5\%.}
}
\end{center}
\end{table}
While beyond LO the difference in the results due to residual scale dependences is below 3\%, at LO a suitable choice of $\mu$ is crucial to minimize the impact of higher order corrections. 
The $K$~factor turns out to be largest for the choice $\mu=m_{W_2}$, where the NLO corrections amount to about 22\% of the LO cross section. For $\mu=Q$, the LO result best approximates the NLO prediction. 
With this setting, the shape of distributions barely changes when going from LO to NLO. This is illustrated by Fig.~\ref{mtnlo},  
\begin{figure}
\begin{center}
\includegraphics[width=7.5cm,height=7.5cm]{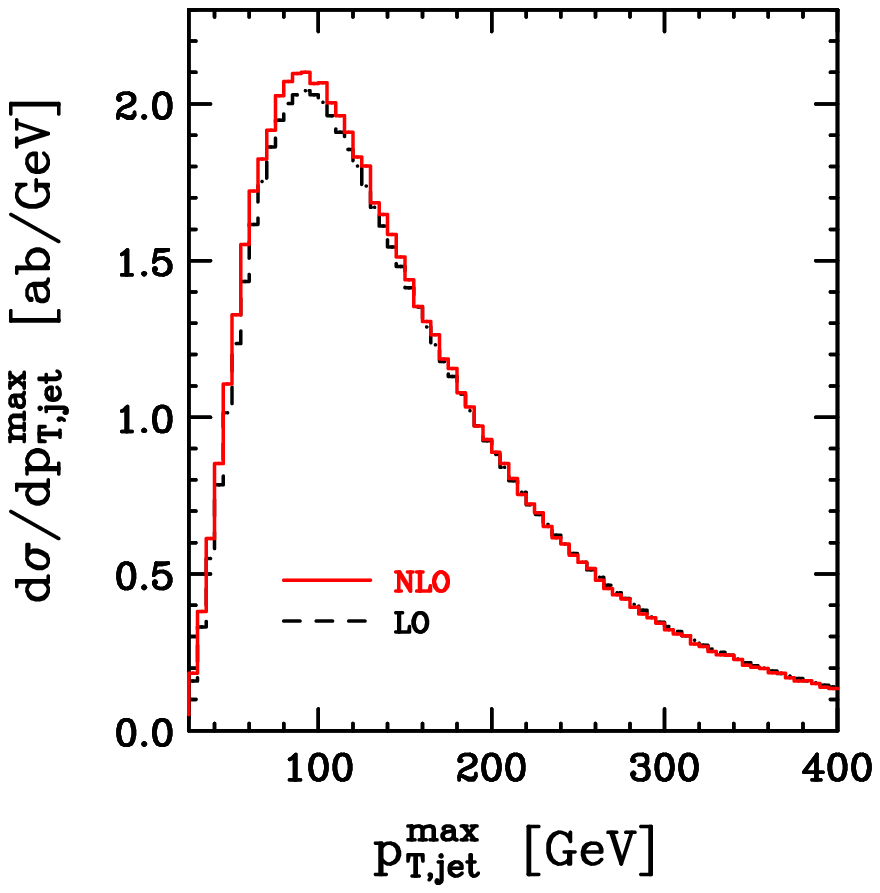}
\hfill
\includegraphics[width=7.5cm,height=7.5cm]{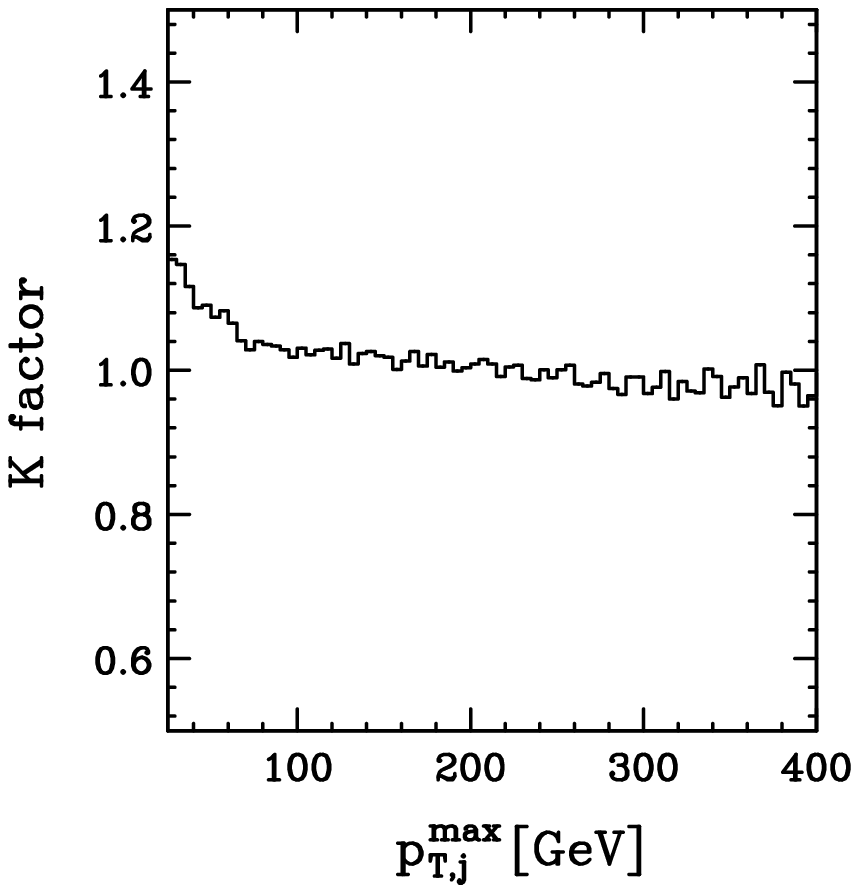}
\caption{\label{ptjnlo}\emph{
LO (black dashed line) and NLO (red solid line) distribution of the tagging jet with the largest transverse momentum the Warped Higgsless scenario A of Tab.~\ref{tab:kk} and differential $K$~factor. Scales are set to 
$\mu_R=\mu_F=Q$.}}
\end{center}
\end{figure}
which shows the transverse cluster mass of the $W^+Z$ system in the Kaluza-Klein scenario at LO and NLO together with the differential $K$~factor. The pronounced resonance behavior of the distribution is retained, with NLO corrections amounting to at most 5\% in the considered range of $m_T(WZ)$. 
Slightly larger shape distortions are found for the transverse momentum distribution of the hardest tagging jet, depicted in Fig.~\ref{ptjnlo}.
At low transverse momenta, NLO corrections of up to $20\%$ are obtained. These are mostly due to the radiation of an extra parton in the real emission contributions, which carries part of the overall transverse momentum available in the reaction. This behavior is completely analogous to the SM case, cf.\ Ref.~\cite{Zeppenfeld:2007ur}.   

In summary, NLO-QCD corrections to VBF processes are as small in Warped Higgsless models as in the SM. With $K$~factors being close to one for cross sections and over a large kinematic range also for distributions, most quantitative estimates can be obtained at the LO level already. For precision predictions, however, the inclusion of NLO-QCD corrections is desirable. 
%
%--------------------------------------------------------------------
%			MAIN SECTION
%--------------------------------------------------------------------
%
\section{Conclusions}
Vector-boson fusion processes represent a promising class of reactions at the LHC: Higgs production via VBF has been discussed as a possible discovery channel of a scalar, spin zero boson as predicted by the SM. Moreover, gauge boson scattering reactions in VBF are sensitive to the mechanism of electroweak symmetry breaking {\em per se}. It is thus vital to access the theoretical uncertainties of weak boson fusion processes at the LHC within the SM and beyond. 

While EW $VVjj$ production in the context of the SM has been studied at NLO-QCD accuracy before \cite{Zeppenfeld:2007ur}, in the present paper we have focused on Warped Higgsless models resulting from the compactification of a gauge theory on an RSI background. In such scenarios, electroweak symmetry breaking is realized by appropriate boundary conditions for the wave functions of the vector bosons along the fifth dimension. The compactification gives rise to towers of Kaluza-Klein gauge bosons, which manifest themselves as high-mass excitations of the photon, the $W^\pm$, and the $Z$ boson in the particle spectrum, and can be interpreted as vector composites from 
the AdS/CFT point of view.  

We have implemented a representative Warped Higgsless model, including NLO-QCD corrections, into the framework of the versatile parton-level Monte Carlo program \textsc{Vbfnlo}, which is publicly available
from \cite{vbfnloweb}. With this code at hand, we have studied the signatures and implications of two particular scenarios. We found that irrespective of the details of the model, cross section enhancements with respect to the SM occur in all production modes, $ZZjj$, $W^+W^-jj$, $W^\pm Zjj$. However, the $W^\pm Zjj$ channel is most sensitive to the Kaluza-Klein excitations $W^\pm_2$, which show up as resonant states in the $W^\pm Z\to W^\pm Z$ sub-amplitudes, yielding characteristic distributions of the decay leptons.
This feature is not obscured by large QCD uncertainties, as an explicit calculation of the dominant NLO-QCD corrections revealed. We have investigated $K$ factors of cross sections and distributions, finding that QCD corrections amount to only a few percent in all kinematic ranges and never exceed about 20\%. Choosing the momentum transfer of the scattering quarks as factorization scale minimizes the effect of the QCD corrections within the Kaluza-Klein scenario. 
This is in complete analogy to what has been found for gauge boson scattering in the SM~\cite{Zeppenfeld:2007ur}. 

%
%--------------------------------------------------------------------
%			MAIN SECTION
%--------------------------------------------------------------------
%
\section*{Acknowledgments}
We would like to thank Manuel B\"ahr, Michael Spannowsky and Malgorzata Worek  for many 
useful discussions.
This research was supported by the Deutsche Forschungsgemeinschaft under SFB TR-9  "Computergest\"utzte Theoretische Teilchenphysik". C.~E.\ was supported by ``KCETA Strukturiertes Promotionskolleg" and B.~J.\ by the Initiative and Networking Fund of the Helmholtz Association, contract HA-101 (``Physics at the Terascale'').
%--------------------------------------------------------------------
%			APPENDIX SECTION
%--------------------------------------------------------------------
\appendix
\section{The Warped Higgsless model}
\label{app:whm}
The compactification of the Warped Higgsless model sketched in Fig.~\ref{fig:hlessmod} explicitly breaks higher-dimensional Lorentz invariance. Yet, the metric defined in Eq.~\gl{rsmetric} is manifestly  Lorentz invariant in four dimensions. Under the unbroken four-dimensional subgroup, a five-dimensional bulk vector field decomposes into a vector and a scalar field in four dimensions. After appropriate bulk gauge-fixing, the scalar components of the gauge fields transform into the longitudinal degrees of freedom of the massive vector bosons in the four-dimensional effective theory \cite{Csaki:2003dt}. 
The wave functions for the $k$th mode of the vector bosons along the fifth dimension are given by solutions of the Bessel differential equation, 
\bee
\label{5dwave}
\psi_k(y)=y \left( a_k J_1(m_k y) + b_k Y_1 (m_k y) \right)\,,
\eee
where $m_k$ is the mass of the Kaluza-Klein state and $k$ is the index of the Sturm-Liouville Problem's solution. The mass spectrum is thus determined by the boundary conditions. Symmetry breaking is triggered by choosing Dirichlet boundary conditions for the gauge fields, while unbroken gauge symmetries are realized via Neumann boundary conditions on the brane.
For the Warped Higgsless Model, the appropriate choice of boundary conditions according to Fig.~\ref{fig:hlessmod} leads to a Kaluza-Klein decomposition of the gauge fields
\bee
\label{eq:kkzdecomp}
Z_\mu^{i}(x,y)&=&\sum_{k} \psi_k^{i;(Z)}(y)Z_\mu^{(k)}(x)\,,\\
\label{eq:kkwdecomp}
W_\mu^{i}(x,y)&=&\sum_{k} \psi_k^{i;(W)}(y)W_\mu^{(k)}(x)\,,
\eee
where $i=L,R$. The mass spectrum of the Kaluza-Klein towers is determined by the solutions of the following set of equations: 
\label{kktower}
\bee
\label{wtower}
W^\pm \hbox{ tower:} & \quad&(R_0-\tilde{R}_1)(\tilde{R}_0-R_1)-(R_0-\tilde{R}_0)(R_1-\tilde{R}_1)=0\\
\label{ztower}
Z \hbox{ tower:} & \quad &
\kappa^2\left\{ (\tilde{R}_0-R_0)(\tilde{R}_1-R_1)+(\tilde{R}_1-R_0)(\tilde{R_0}-R_1)\right\}\nonumber \\
& &\hspace{3cm}+ 2 (\tilde{R}_1-R_0)(\tilde{R}_0-R_1)=0 \\
\label{z'tower}
 \hbox{photon tower:} &\quad &\tilde{R}_0-R_0=0
\eee
with 
\bee\label{rrtdef}
\kappa = {g_5\over \tilde{g}_5}\,, \qquad
R_i={Y_i(m_kR)\over J_i(m_kR)}\,, \quad 
\tilde{R}_i={Y_i(m_kR')\over J_i(m_kR')}\,.
\eee 
Inserting the Kaluza-Klein-decomposition, Eqs.~\gl{eq:kkzdecomp}-\gl{eq:kkwdecomp}, into the Lagrangian of the five-dimensional theory and integrating over the compactification direction $y$, the gauge boson couplings can be expressed in terms of their eigenfunctions \gl{5dwave}, yielding, e.~g., 
\bee
\label{3coup}
g_{W_kW_1Z_1}&=&g_5^2\int_R^{R'} \d y \, {R\over y}\sum_{i=L,R} \left\{ \psi^{i;(W)}_k \psi^{i;(W)}_1 \psi^{i;(Z)}_1 \right\}\,.
\eee
The sum rules, Eqs.~\gl{sumww1}-\gl{sumwz2}, follow from a completeness relation for the wave functions along the $y$ direction and the underlying higher-dimensional gauge invariance \cite{Csaki:2003dt}.

A code that determines the KK masses and couplings needed for the results of Secs.~\ref{sec:lhcpred} 
and \ref{sec:qcdcor} is built into \textsc{Vbfnlo}.

%--------------------------------------------------------------------
%			BIBLIOGRAPHY
%--------------------------------------------------------------------

\end{document}